\documentclass[pre,aps,showpacs,preprint]{revtex4}

\usepackage{amsfonts,amssymb,amsmath,graphicx,epsfig,amscd,ulem}
\usepackage{graphicx}
\usepackage{array}

\def\bi{\begin{itemize}}

\def\ei{\end{itemize}}
\def\be{\begin{equation}}
\def\ee{\end{equation}}
\def\bea{\begin{eqnarray}}
\def\eea{\end{eqnarray}}

\def\gdot{\dot\gamma}
\def\taut{\tau_{transition}}
\def\taua{\tau_{aging}}
\def\taur{\tau_{\scriptscriptstyle 0}}

\def\tt{t_{transition}}
\def\ta{t_{aging}}

\newcommand{\dnp}{n_{\text a}}
\newcommand{\dnm}{n_{\text r}}
\newcommand{\fp}{f_{\text a}}
\newcommand{\fm}{f_{\text r}}
\newcommand{\lp}{\ell_{\text a}}
\newcommand{\lm}{\ell_{\text r}}
\newcommand{\km}{k_{\text r}}
\newcommand{\kp}{k_{\text a}}
\newcommand{\pp}{p_{\text a}(\theta)}
\renewcommand{\pm}{p_{\text r}(\theta)}
\newcommand{\ppo}{p_{\text a}^{o}(\theta)}
\newcommand{\pmo}{p_{\text r}^{o}(\theta)}

\DeclareTextSymbol{\degre}{OT1}{23}

\begin{document}

\author{Guillaume Ovarlez\footnote{corresponding author: guillaume.ovarlez@lcpc.fr}, Xavier
Chateau}
\affiliation{Universit\'e Paris Est - Institut Navier\\
Laboratoire des Mat\'eriaux et Structures du G\'enie Civil (LCPC-ENPC-CNRS)\\
2, all\'ee Kepler, 77420 Champs-sur-Marne, France}

\title{Influence of the shear stress applied during the flow stoppage and the rest on the mechanical properties of thixotropic suspensions}

%\date{}

\begin{abstract}
We study the solid mechanical properties of several thixotropic
suspensions as a function of the shear stress history applied
during their flow stoppage and their aging in their solid state.
We show that their elastic modulus and yield stress depend
strongly on the shear stress applied during their solid-liquid
transition (i.e. during flow stoppage) while applying the same
stress only before or only after this transition may induce only
second-order effects: there is negligible dependence of the
mechanical properties on the preshear history and on the shear
stress applied at rest. We also found that the suspensions age
with a structuration rate that hardly depends on the stress
history. We propose a physical sketch based on the freezing of a
microstructure whose anisotropy depends on the stress applied
during the liquid/solid transition to explain why the mechanical
properties depend strongly on this stress. This sketch points out
the role of the internal forces in the colloidal suspensions
behavior. We finally discuss briefly the macroscopic consequences
of this new phenomenon and show the importance of using a
controlled-stress rheometer.
\end{abstract}
\maketitle

\section{Introduction}\label{section_introduction}

Dense suspensions arising in industrial processes (concrete
casting, drilling muds...) and natural phenomena (debris-flows...)
often involve a broad range of particle sizes. The behavior of
these materials reveal many complex features which are far from
being understood (for a recent review, see \cite{Stickel2005}).
This complexity originates from the great variety of interactions
between the particles (colloidal, hydrodynamic, frictional,
collisional...) and of physical properties of the particles
(volume fraction, deformability, sensitivity to thermal agitation,
shape, buoyancy...) involved in their behavior.

Basically, these materials exhibit a yield stress and have a solid
viscoelastic behavior below this yield stress; above the yield
stress they behave as liquids: they flow. This yielding behavior
originates from the colloidal interactions which create a jammed
network of interacting particles \cite{Larson1999,Coussot2005}.
These materials also exhibit thixotropic behaviors: their time
dependent properties and the characteristic time to reach a steady
state flow depend on the previous flow history
\cite{mewis1979,Coussot2005,dullaert2005}. Considerable work has
been devoted to studying the structure of suspensions under
stationary shear flow and its link to the rheological properties
of the suspensions (see e.g.
\cite{Larson1999,Vermant2005,Stickel2005} for a recent review). It
has been well established that the changes in the material
properties as a function of the shear history are linked to
structural changes. Moreover, when these materials are left at
rest (the rest is usually defined as a period characterized by a
naught shear rate), their static yield stress $\tau_c$ (the shear
stress one has to impose to start a flow) increases with the
resting time \cite{Cheng1986,Coussot2002,derec2003}, and may be
one order higher than their dynamic yield stress $\tau_d$ (the
shear stress for flow cessation). In parallel, the elastic modulus
is also found to increase with the resting time
\cite{derec2003,manley2005,Coussot2006}. These features are shared
by many aggregating suspensions and colloidal glasses: in the case
of aggregating suspensions, the evolution of the behavior may be
explained by a reversible decrease of the flocculation state under
shear or an increase if the material is left at rest
\cite{pignon1998}. In the case of colloidal glasses, the evolution
is related to the evolution of the microstructure through a
cage-diffusion process \cite{abou2001,bonn1999}. In both cases,
the evolution at rest is related to Brownian motion of the
particles. It is worth noting that physical phenomena that do not
occur at the particle scale may also produce mechanical aging.
Recently, \citet{manley2005} proposed that the mechanical aging of
their aggregating suspensions, which is observed without any
dynamics at the particle scale, may be attributed to the increase
of the contact area between the particles in time.

Applying a given stationary preshear to a thixotropic suspension
should enable to obtain a uniquely defined structure of the
material: this structure should depend only on the value of the
shear rate applied during the preshear. Then, once the flow is
stopped after this given preshear, the mechanical characteristics
(elastic modulus and yield stress) of the suspension in its solid
state and their evolution at rest would be uniquely determined.
However, thixotropic materials may pass from a liquid to a solid
state for any stress history below a well defined dynamic yield
stress \cite{Coussot2006}. As far as we know, the effect of the
stress history (below the dynamic yield stress) applied during the
unsteady flow leading to structural build-up on the structure and
the mechanical properties of thixotropic suspensions has not been
studied. Moreover, the effect of this stress on the aging of yield
stress materials has only been poorly studied.

Recently, \citet{Ovarlez2007} have shown that when a stress lower
that the dynamic yield stress is applied after a strong preshear
to a thixotropic suspension in a fully destructured liquid state,
its flow stoppage (i.e. its liquid/solid transition) is delayed by
a time that increases with the applied stress and diverges at the
approach of the dynamic yield stress. They also observed that the
elastic modulus of the material once in its solid regime depends
on the small stress (below the dynamic yield stress) that is
applied after the preshear. However, they did not focus on this
feature.

\citet{Cloitre2000} have observed that the aging of microgel
pastes, as probed by the creep response to a shear stress below
the yield stress, gets slower when the applied shear stress is
increased. This phenomenon was interpreted as resulting from a
competition between aging and partial rejuvenation induced by
large rearrangements when the shear stress exceeds a value
corresponding to the end of the linear regime of the material. It
has also been shown by \citet{Viasnoff2002} that the picture may
be more complex. They studied the aging of a colloidal suspension
through the evolution in time of the microscopic rearrangement
kinetics of the particles, probed by DWS. They found that applying
oscillations (below the yield strain) to a colloidal suspension at
rest may induce either an overaging or a rejuvenation (depending
on the strain). Mechanical aging has been observed in frictional
materials: the static coefficient of friction increases with the
time of rest \cite{Berthoud1999}. This aging, i.e. the increase
rate of the static friction coefficient, was found to be strongly
accelerated when a stress is applied during rest in solid on solid
friction experiments \cite{Berthoud1999} and in granular materials
\cite{Losert2000,Restagno2002,Ovarlez2003}. Direct frictional
contacts between particles have been proposed to be at the origin
of the yield stress \cite{Furst2007} and of the mechanical aging
\cite{manley2005} of colloidal gels; in this framework, we would
thus expect the mechanical aging of such suspensions to be
strongly accelerated under stress.

In this paper, we question the influence of the shear stress
history under the dynamic yield stress on the solid mechanical
properties of thixotropic materials. We control accurately the
shear stress imposed during the flow stoppage and the rest of
various thixotropic suspensions, and measure their yield stress
and elastic modulus evolution in time. We show that their solid
mechanical properties depend strongly on the stress applied during
their liquid/solid transition (i.e. their flow stoppage), while
the stresses applied only before the liquid/solid transition (i.e.
during the flow in the liquid state) or only after the
liquid/solid transition (i.e. during the aging at rest) may induce
only second-order effects. We also show that the suspensions age
with a structuration rate that hardly depends on the stress
history. The materials and methods used to perform this study are
presented in Sec.~\ref{section_display}. The effect of the stress
history on the overall mechanical behavior and on the aging is
shown in Sec.~\ref{section_mechanics}. We discuss the physical
origin of the observed behaviors and present some important
macroscopic consequences in Sec.~\ref{section_discussion}.

\section{Materials and methods}\label{section_display}

\subsection{Pastes}\label{section_materials}

We perform most experiments on bentonite suspensions. In order to
check the generality of our results, we also studied three other
thixotropic materials and a simple yield stress fluid: a mustard,
a silica suspension, a thixotropic emulsion and a simple emulsion.

Bentonite suspensions are made of (smectite) clay particles of
length of order 1$\mu$m and thickness 10nm \cite{luckham1999}. The
particle can aggregate via edge-to-face links, so that the
suspension may be seen as a colloidal gel \cite{luckham1999}. As a
consequence, these suspensions are thixotropic. Moreover, at rest,
their yield stress and elastic modulus increase in time. The
mechanical properties may be varied by varying the particle volume
fraction: we prepare 3 suspensions, at 6, 9, and 10\% volume
fraction, that have an initial (i.e. 100s after the end of a
preshear at high shear rate) static yield stress of 30, 50 and
65Pa. Each sample was prepared by a strong mixing of the solid
phase with water. Then, the suspensions were left at rest for 3
months before any test, which avoids further irreversible
(chemical) aging over the duration of the experiments.

The mustard (Maille, France) is a mixture of water, vinegar,
mustard seed particles, mustard oil, and various acids. We may see
it as a suspension in an oil-in-water emulsion with a large
concentration of elements (droplets and particles).

The silica suspension is a suspension of silica particles (Rhodia)
of 3.7$\mu$m average diameter at a 22\% volume fraction; KCl is
added at a 0.3M concentration in order to destabilize the
suspension and form a concentrated aggregating suspension.

As a simple emulsion and a thixotropic emulsion, we use the
materials of \citet{Ragouilliaux2007}. The pure emulsion is
prepared by progressively adding water in an oil-surfactant
solution (Sorbitan monooleate, 2\%) under high shear. The water
droplets have a 1$\mu m$ diameter. Their concentration is fixed at
70\%, which means that the droplets are in close contact. This
simple emulsion behaves as a simple yield stress fluid (i.e. it
exhibits no time dependent behavior). A thixotropic emulsion is
made by loading the simple emulsion with colloidal particles
\cite{Ragouilliaux2007}. The surfactant and the colloidal
particles (hydrophobic clay particles (Bentone 38, Elementis
Specialties company) with a mean diameter 1$\mu$m and thickness 10
nm) are first mixed at a solid volume fraction of 3\% in the oil;
the water is then progressively added in this suspension under
high shear. As shown by \citet{Ragouilliaux2007}, this loaded
emulsion has a thixotropic behavior. However, this behavior does
not originate from the colloidal interactions between the clay
particles: the bentonite suspension in oil is a simple fluid. It
is thus likely that the clay particles tend to form links between
neighboring droplets: the dynamics of these links formation is at
the origin of the thixotropic behavior of the suspension loaded
with colloidal particles.

Finally, we have prepared a wide range of materials, from a low
volume fraction suspension which may form a loose fractal gel to a
dense emulsion in which all droplets are in close contact. The
thixotropic and aging behaviors of these materials may have
various origins such as aggregation, new links creation between
particles in close contact, and rearrangements of the particles
configuration. By performing our experiments on all these
materials, we will thus be able to determine whether the behaviors
we observe are generic properties shared by thixotropic
suspensions, or if they are specific to a given material or a
given physical mechanism.

\subsection{Stress histories and rheometrical measurements}\label{section_rheom}

Most rheometric experiments are performed within a coaxial
cylinder thin gap Couette geometry (inner radius $R_i=17.5$mm,
outer cylinder radius $R_e =18.5$mm, height $H=45$mm) on a
commercial rheometer (Bohlin C-VOR 200) that imposes either the
torque or the rotational velocity (with a torque feedback). This
ensures having a roughly homogeneous stress in the gap. We also
checked in a cone and plate geometry (4$\degre$\!, radius 2cm)
that we observe the same phenomena as in the thin gap Couette
geometry. In order to avoid wall slip \cite{Coussot2005}, we use
sandblasted surfaces of roughness larger than the size of the
particles.

\subsubsection*{\it Procedures}

As the materials we study are thixotropic, it is necessary to
strongly preshear the materials after loading in order to perform
all the measurements in the same conditions, i.e. to start always
from a same state of structuration of a material. In all
experiments, we thus first preshear the material at high stresses
(corresponding to a high shear rate of about 100s$^{-1}$) during
200s in order to start from a fully destructured state of the
material in its liquid regime.

\begin{figure}[htbp!] \begin{center}
\includegraphics[width=11cm]{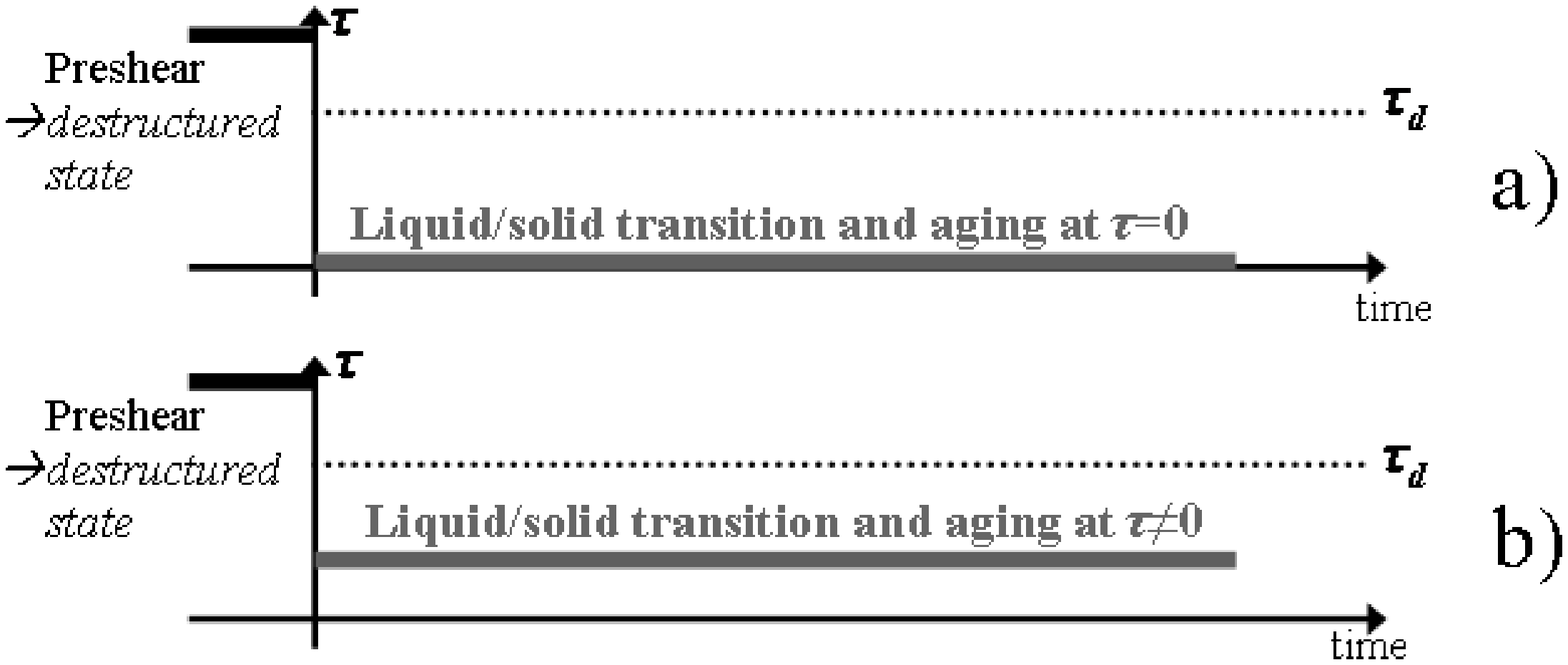}
\includegraphics[width=11cm]{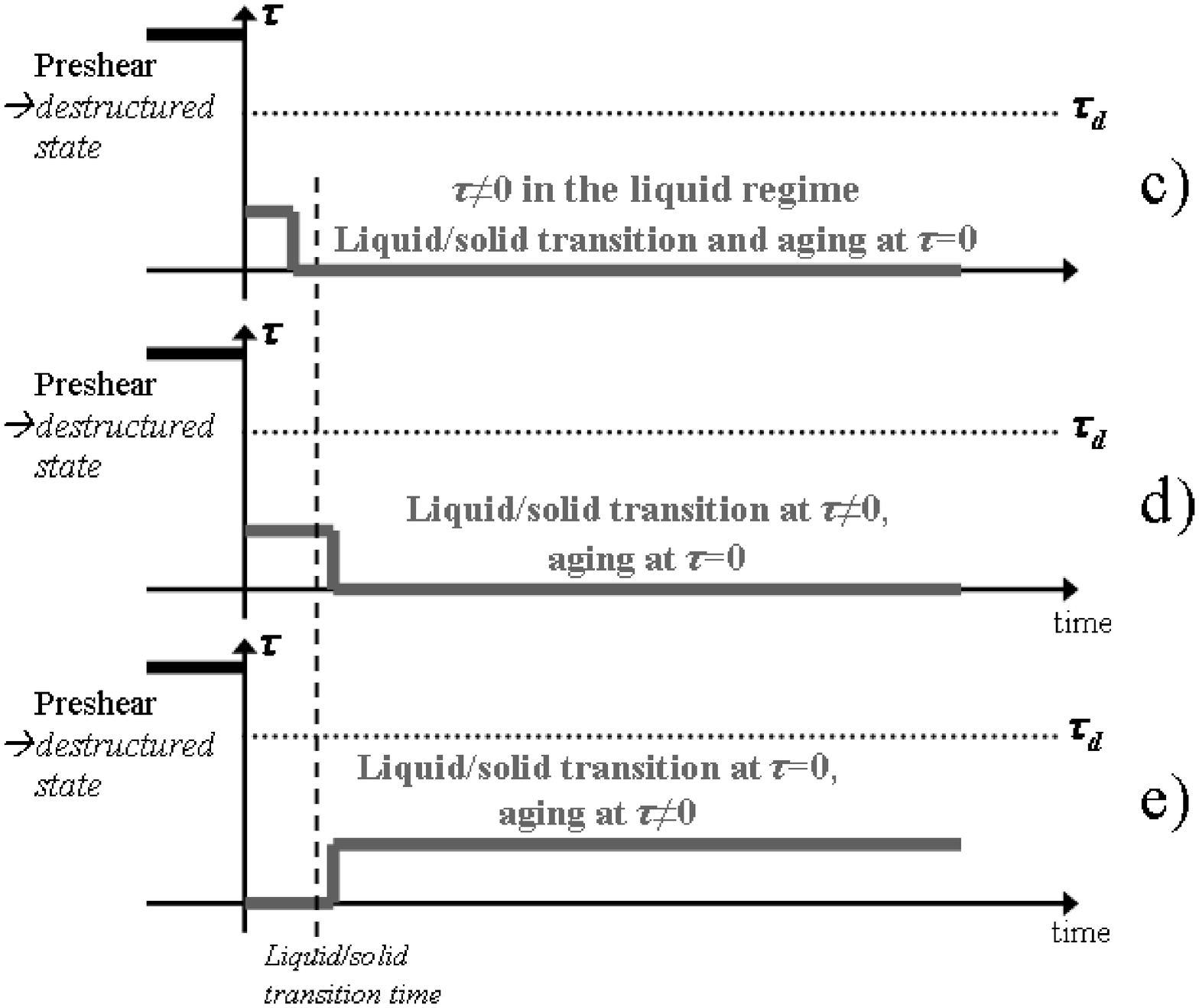}
\caption{Shear histories. In all experiments, a strong preshear is
first applied to the material in its liquid state. Then: a) a
naught shear stress is applied ($\taur=0$Pa); b) a stress
$\taur\neq0$Pa lower than the dynamic yield stress $\tau_d$ is
applied during the liquid/solid transition and all the aging; c) a
stress $\taur\neq0$Pa lower than the dynamic yield stress $\tau_d$
is applied during flow deceleration as long as the material is in
a liquid state, then the stress is lowered to zero during the
liquid/solid transition and the aging; d) a stress $\taut\neq0$Pa
lower than the dynamic yield stress $\tau_d$ is applied during the
material liquid/solid transition and then the stress is lowered to
zero during the aging; e) a naught stress is applied during the
liquid/solid transition and then a stress $\taua\neq0$ lower than
the dynamic yield stress $\tau_d$ is applied during the aging in
the solid regime.}\label{fig_procedures}
\end{center} \end{figure}

After the preshear, we impose various shear stress histories
$\taur(t)$ below the dynamic yield stress $\tau_d$: the materials
may then stop flowing and age at rest under various stresses. Only
simple histories, in which the stress is piecewise-constant will
be considered; the exact procedures will be detailed below (see
Fig.~\ref{fig_procedures}). In order to study the influence of
this shear stress history on the mechanical properties at rest of
the materials, we measure the evolution in time of their elastic
modulus. We also measure their yield stress as a function of the
duration of the stress history $\taur(t)$. Before performing the
yield stress measurement, we first relax the elastic strain by
lowering the stress to 0Pa during 5 seconds.

\subsubsection*{\it Elastic modulus measurements}

The evolution in time of the suspensions elastic modulus is
measured by superposing small stress oscillations to the stress
history $\taur(t)$. We impose
$\tau(t)=\taur(t)+\delta\tau_{\scriptscriptstyle
0}\cos(\omega_{\scriptscriptstyle 0}*t)$ and we measure the strain
response
$\gamma(t)=\gamma_{creep}(t)+\delta\gamma_{\scriptscriptstyle
0}(t)\cos(\omega_{\scriptscriptstyle 0}*t+\phi(t))$. The elastic
modulus is then simply $G(t)=\delta\tau_{\scriptscriptstyle
0}/\delta\gamma_{\scriptscriptstyle 0}(t)$.

In most experiments, the oscillatory shear stress is applied at a
frequency of 1Hz. The amplitude $\delta\tau_{\scriptscriptstyle
0}$ depends on the sample: it is chosen so as to ensure that all
materials are tested in their linear regime (the strain is always
lower than $10^{-3}$). These experiments were performed with
several different amplitudes on some materials in order to check
that the results are independent of the choice of
$\delta\tau_{\scriptscriptstyle 0}$. We also checked the
independence of the results on the frequency; although the overall
value of the elastic modulus depends slightly on the frequency,
the aging rate and the effect of the stresses histories on the
mechanical properties we evidence in this paper are not sensitive
to the frequency. Finally, we checked that such oscillations do
not affect the mechanical properties of the materials: the same
elastic modulus is measured after a long time whether oscillations
are applied or not during this time. This may seem in
contradiction with the observations of \citet{Viasnoff2002}, who
found that there is an interplay between the aging and the
oscillations, but apart from the differences between their system
and ours, it must be noted that they imposed oscillatory strains
higher than 2.9\% whereas we impose oscillatory strains lower than
0.1\%. Actually, from the results of \citet{Cloitre2000}, it seems
that probing the material in its linear regime prevents aging from
being affected by the oscillations. Note also that as we measure
the linearized response of the material around a stress that may
be non-zero, we have to check that differences between the elastic
moduli measured for various shear stresses histories are not due
to non linear effects. This will be shown in
Sec.~\ref{section_elasticity}. Note finally that there is a small
creep $\gamma_{creep}(t)$ in response to the applied stress
history. As the measurements are performed in the solid regime of
the materials, the creep flow is continuously decelerating
\cite{Coussot2006}. In all the experiments we performed, except
during the first few seconds after imposing $\taur$, the creep
rate $\gdot_{creep}(t)$ was always $<10^{-4}s^{-1}$ so that there
was no interplay between the negligible creep and the elasticity
measurement.

\subsubsection*{\it Yield stress measurements}

\begin{figure}[htbp] \begin{center}
\includegraphics[width=7cm]{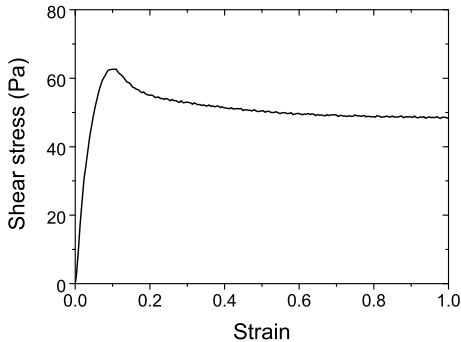}
\caption{Shear stress vs.\ strain when slowly shearing a 9\%
bentonite suspension from rest at $10^{-2}$s$^{-1}$ 600s after the
end of the preshear.}\label{fig_yield_measurement}
\end{center} \end{figure}

We mainly perform our yield stress measurements by means of a
velocity controlled method \cite{nguyen1985}: the inner cylinder
is driven at a low velocity, and the yield stress is defined by
the overshoot presented by the shear stress in a shear stress vs.\
strain plot (see an example on Fig.~\ref{fig_yield_measurement}
for a 9\% bentonite suspension). This yields a good evaluation of
the yield stress as the overshoot is followed by a slow stress
decrease: this means that the material starts to be destructured
at the overshoot (as long as the shear timescale is lower than the
structuration timescale) and thus flows. Note that, as the
materials we study are thixotropic, the value of the yield stress
depends on the time elapsed between the end of the preshear and
the measurement: this resting time must then be controlled. We
chose to drive the inner cylinder in order to induce a shear rate
of 0.01s$^{-1}$. We checked that the features we observe in this
paper do not depend on the low shear rate that is imposed to
measure the yield stress. We finally checked that our results are
independent of the measurement procedure: we also performed linear
shear stress ramps \cite{uhlherr2005}; in these experiments the
shear stress is increased linearly in time and is plotted vs.\ the
shear rate in order to identify the value of the shear stress at
the onset of flow \cite{Ovarlez2007b}.

\section{Experimental results}\label{section_mechanics}

\begin{figure}[htbp] \begin{center}
\includegraphics[width=7.9cm]{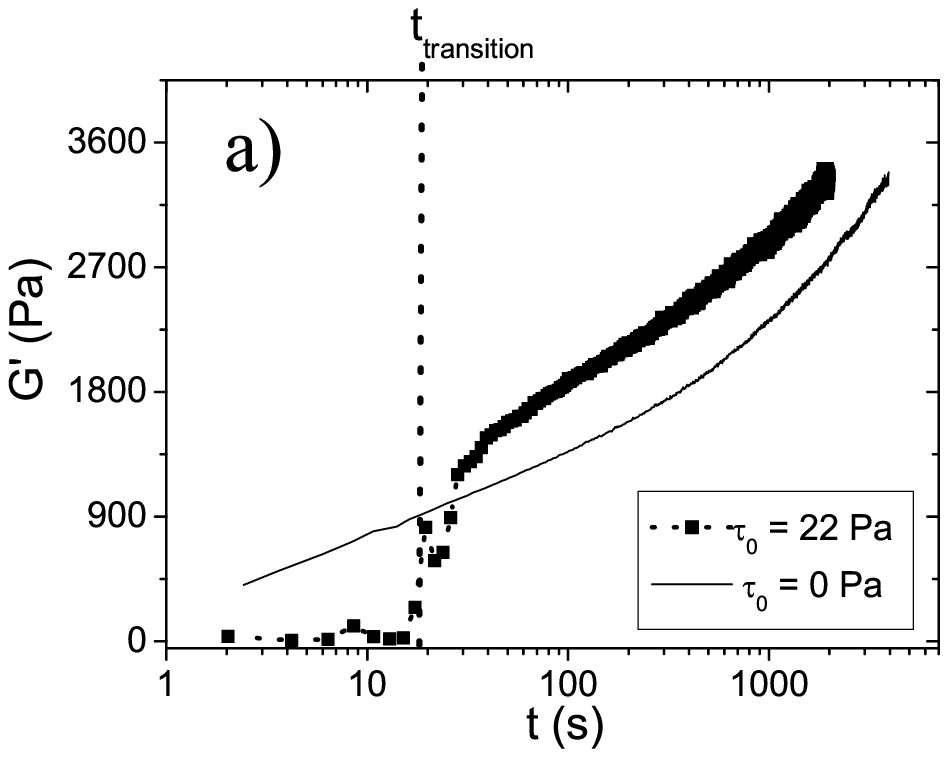}
\includegraphics[width=7.2cm]{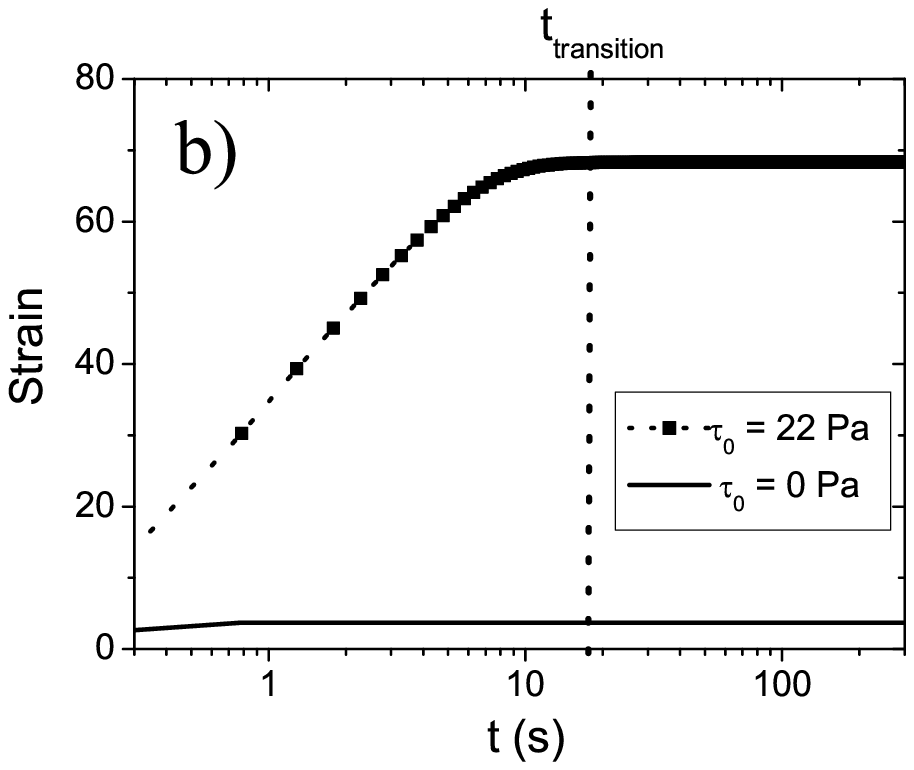}
\caption{a) Elastic modulus G' vs.\ time $t$ after strongly
shearing a 9\% bentonite suspension, for two different stresses
applied after the preshear (see Fig.~\ref{fig_procedures}a,b):
$\taur=0$Pa (line) and 22Pa (squares). b) Strain vs.\ time after
strongly shearing a 9\% bentonite suspension, for the same
experiments as in Fig.~\ref{Fig_effet_cont_sur_elas1}a (by
convention the strain is chosen as 0 at $t=0$s). The vertical
dotted line delimits the liquid and solid regimes of the bentonite
suspension in the experiment performed at $\taur=22$Pa, and
defines the liquid/solid transition time
$\tt$.}\label{Fig_effet_cont_sur_elas1}
\end{center} \end{figure}

Let us first show the impact of applying a constant stress $\taur$
below the dynamic yield stress $\tau_d$ to a thixotropic
suspension after a preshear (procedures of
Fig.~\ref{fig_procedures}a,b). In
Fig.~\ref{Fig_effet_cont_sur_elas1}a we plot the elastic modulus
of a 9\% bentonite suspension vs.\ the time elapsed since the end
of the preshear, for $\taur=0$Pa and $\taur=22$Pa. In both cases,
the elastic modulus G' increases in time. This is characteristic
of the aging at rest of thixotropic materials
\cite{derec2003,Coussot2006}. However, depending on the applied
shear stress $\taur$, we observe striking differences in the
behavior at short times. While the elastic modulus G' increases
regularly from time $t=0$s when a zero shear stress is applied
after the preshear, when a non-zero shear stress $\taur$ is
applied, G' is equal to zero (within the measurement uncertainty)
in a first stage and it suddenly starts increasing at some time
$\tt\neq 0$ (here equal to 19s for $\taur$=22Pa). Afterwards, we
see that both moduli increase regularly in time, with roughly the
same increase rate. However, at a given time, the modulus measured
when a stress $\taur\neq 0$ is applied is much higher than the
modulus observed at zero stress: here, for $\taur=22$Pa, G' is
600Pa higher than for $\taur$=0Pa. As G' is measured as the
linearized response of the material to oscillations around
different $\taur$ in these experiments, we checked that
differences between the elastic moduli measured with both
procedures are not due to non linear effects: we showed that the
same modulus is measured around $\taur\neq 0$ as the one measured
around 0Pa when removing the stress $\taur$ after a long time.

In the following, we first show that the sudden increase in time
of the elastic modulus we observe for $\taur\neq0$ is the
signature of a liquid/solid transition. We then show that the
large difference in the elastic modulus values observed for
different $\taur$ is due mainly to the stress applied during this
liquid/solid transition. We also show that the stress applied only
during the flow before stoppage and during the aging in the solid
state have basically no influence on the mechanical properties. We
finally quantify the link between the applied stress at the liquid
solid transition and the increase of the elastic modulus and the
yield stress of the materials. We study the behavior of the
bentonite suspensions in detail and show that these features are
shared by all the thixotropic materials we studied.

\subsection{Liquid/solid transition}\label{section_elasticity}

As shown in detail by \citet{Ovarlez2007}, the sudden appearance
of an elastic behavior in colloidal suspensions observed in
Fig.~\ref{Fig_effet_cont_sur_elas1} is the signature of a
liquid/solid transition of the material. Superposition of
oscillations to a constant shear stress actually allows to
identify unambiguously the liquid and the solid states of
thixotropic materials \cite{Ovarlez2007}. The constant stress
probes the flow properties, while the superimposed small
oscillations probe the actual material strength. In
Fig.~\ref{Fig_effet_cont_sur_elas1}b we plot the strain response
to the constant stress vs.\ time after the same stress step as in
the experiments of Fig.~\ref{Fig_effet_cont_sur_elas1}a; by
convention the strain is chosen as 0 at $t=0$s. We observe that,
when a stress $\taur$=0Pa is applied, the flow stops within a few
100ms (due to fluid inertia) corresponding to a strain of around
3.7, whereas when a stress $\taur$=22Pa is applied, the strain
increases during around 20s, resulting in a strain of order 70;
afterwards, the strain saturates. Consistently, the beginning of
the plateau of strain corresponds to the sudden appearance of a
substantial elastic modulus in the material. This indicates that
the material is now in a solid state, while it was in a liquid
state during its flow. Finally, these observations allow to
identify precisely and unambiguously the liquid regime (the
material flows and has a negligible elastic modulus) and the solid
regime (the material stops flowing and gets a substantial elastic
modulus) of the material. As shown by \citet{Ovarlez2007}, this
identification of the liquid and solid regimes is consistent with
the loss modulus G'' measurements: G'' is larger than the elastic
modulus G' and is proportional to the apparent viscosity of the
material during the flow; at the liquid/solid transition, G''
starts to decrease while G' abruptly increases and crosses over
the G'' curve at its peak. The liquid/solid transition occurs
after a time $\tt$ (equal to 19s in
Fig.~\ref{Fig_effet_cont_sur_elas1}) that increases with $\taur$,
and which tends to infinity when $\taur$ tends to a stress
$\tau_d$ (see \cite{Ovarlez2007}), i.e. the material remains
indefinitely in a liquid state for $\taur>\tau_d$: this defines
precisely $\tau_d$ as the dynamic yield stress.

\begin{figure}[htbp]
\begin{center}
\includegraphics[width=7.9cm]{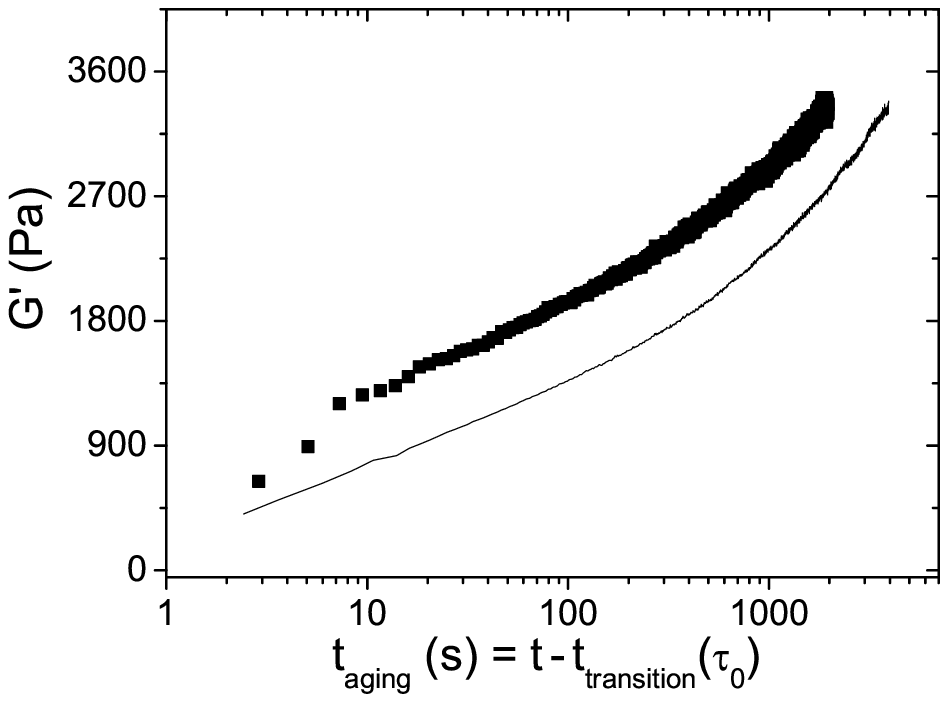}
\caption{Same data as in Fig.~\ref{Fig_effet_cont_sur_elas1}a
plotted vs.\ the time spent in the solid regime
$\ta=t-\tt(\taur)$.}\label{Fig_effet_cont_sur_elas1b}
\end{center}
\end{figure}

In the following, we thus redefine the relevant aging time as the
time spent by the material in its solid regime:
$\ta=t-\tt(\taur)$. The G' data of
Fig.~\ref{Fig_effet_cont_sur_elas1}a are replotted vs.\ $\ta$ in
Fig.~\ref{Fig_effet_cont_sur_elas1b}. Both elastic moduli now seem
to increase at the same rate during the whole aging time and to
differ basically only by a constant value $\Delta G'$ (here around
600Pa) that may depend on $\taur$. We investigate the origin of
these moduli differences on the stress history in the following.

\subsection{Influence of the stress history}

It is known that the structure and the steady state flow
properties of suspensions depend on their shear history. It is
thus first worth wondering if the effect we observe is the
consequence of a dependence on the applied stress of the
structural build-up in the liquid regime before flow stoppage.
Moreover, as stated in the introduction, a stress applied on a
solid aging material may be expected to change its behavior. We
can then also wonder if the effect we observe is the consequence
of a dependence on the applied stress of the structural evolution
of the material. That is why we study the impact on the mechanical
properties of the stress applied during these three phases: the
shear flow in the liquid regime, the liquid/solid transition, and
the aging in the solid regime. In order to separate the effects of
the applied stress in these three phases, we have applied the
following stress histories (see Fig.~\ref{fig_procedures}c,d,e),
in complement to the simple stress histories of
Fig.~\ref{fig_procedures}a,b:
\begin{itemize}
\item in a first series of experiments
(Fig.~\ref{fig_procedures}c), after a preshear, we apply a
non-zero stress $\taur<\tau_d$ during a time
$t_{\scriptscriptstyle 0}<\tt(\taur)$ just too short for the
liquid/solid transition to occur, before relaxing the stress to
zero during the liquid/solid transition and the aging at rest; we
then measure the elastic modulus and yield stress evolution in
time. These experiments test the influence of the stress applied
in the liquid regime. \item in a second series of experiments
(Fig.~\ref{fig_procedures}d), just after the preshear we impose a
non-zero stress $\taut<\tau_d$ during a time sufficient for the
material to stop flowing and pass from a liquid to a solid state
under this stress. Afterwards, we apply no stress during the aging
($\taua=0$Pa) and we measure the elastic modulus and yield stress
evolution in time. When compared with the experiment of
Fig.~\ref{fig_procedures}c, these experiments test the influence
of the stress applied during the liquid/solid transition.\item in
a third series of experiments (Fig.~\ref{fig_procedures}e), just
after the preshear we first impose a naught stress during a time
$t_{stop}=$30s sufficient for the material to pass from a liquid
to a solid state and then we impose a non-zero stress
$\taua<\tau_d$ during the aging in the solid state; we then
measure the elastic modulus and yield stress evolution in time.
These experiments test the influence of the stress applied during
the aging in the solid state.
\end{itemize}

\begin{figure}[htbp] \begin{center}
\includegraphics[width=7.9cm]{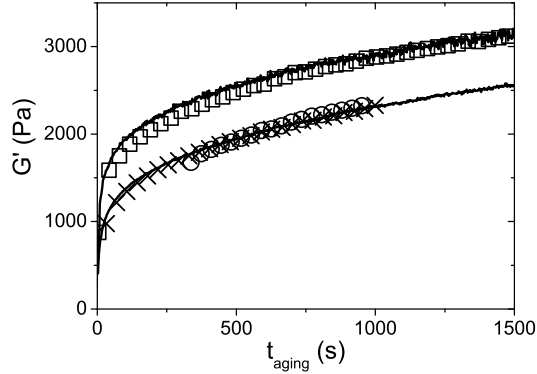}
\caption{Elastic modulus G' vs.\ time after strongly shearing a
9\% bentonite suspension, measured with the various procedures of
Fig.~\ref{fig_procedures}: a) when a stress $\taur=0$Pa is applied
during the whole experiment (lower line); b) when a stress
$\taur=22$Pa is applied during the whole experiment (upper line);
c) when a stress $\taur=22$Pa is applied during
$t_{\scriptscriptstyle 0}=14$s ($t_{\scriptscriptstyle 0}<\tt$)
and then lowered to zero during the liquid/solid transition and
the aging (crosses); d) when a liquid/solid transition stress
$\taut=22$Pa is applied during $t_{\scriptscriptstyle 0}=22$s
($t_{\scriptscriptstyle 0}>\tt$) and then lowered to zero during
the aging (empty squares); e) when a liquid/solid transition
stress $\taut=0$Pa is applied during 30s and then an aging stress
$\taua=22$Pa is applied during 300s (empty
circles).}\label{Fig_effet_cont_sur_elas2}
\end{center} \end{figure}

The elastic modulus measurements performed on the 9\% bentonite
suspension with the procedures of
Fig.~\ref{fig_procedures}a,b,c,d,e are plotted vs.\ time in
Fig.~\ref{Fig_effet_cont_sur_elas2}.

\subsubsection*{\textbf{\textit{Impact of the stress applied in the liquid
regime}}}

Importantly, with the procedure of Fig.\ref{fig_procedures}c, we
found that when applying $\taur\neq0$ only during a time too short
for the liquid/solid transition to occur (here this time was
chosen as 14s), the elastic modulus value and its evolution in
time are the same as when $\taur=0$Pa during the whole experiment
(Fig.~\ref{Fig_effet_cont_sur_elas2}). The conclusion is that
there is no significant dependence of the elastic modulus on the
shear history in the liquid regime before flow stoppage. We
checked this feature for several values of $\taur$. This means
that while the intensity of the shear may probably influence the
material state, as shown by many authors for steady-state flows
\cite{Vermant2005}, it is clearly not at the origin of the strong
elastic modulus strengthening we evidence here (note that the
shear rate value just before lowering $\taur$ to 0Pa in
Fig.~\ref{Fig_effet_cont_sur_elas2} was of order 0.1s$^{-1}$
whereas its is of order 100s$^{-1}$ during the preshear). Compared
with what we observe, the shear history in the liquid regime may
induce only second-order effects on the solid mechanical
properties of the suspensions.

We also checked that the intensity of the preshear applied before
lowering the stress below $\tau_d$ has no influence on the
results. Finally, we have performed experiments in which the
stress $\taut$ is applied in the direction opposite to the
preshear. The results, presented in
Fig.~\ref{Fig_effet_cont_sur_elas_bento1}a, show that the same
result is obtained whatever the relative direction of the preshear
and of $\taut$. This means that, while it is known that a preshear
in a given direction creates an anisotropic microstructure
\cite{mathis1988,parsi1987}, the impact of this anisotropy on the
solid mechanical properties is negligible or erased by the shear
stress history applied after the preshear.

\subsubsection*{\textbf{\textit{Impact of the stress applied during the
liquid/solid transition}}}

In order to evaluate the impact of the stress applied during the
liquid/solid transition of the material, we plot in
Fig.~\ref{Fig_effet_cont_sur_elas2} the elastic modulus vs.\ time
for the procedure depicted on Fig.~\ref{fig_procedures}d applied
to the 9\% bentonite suspension: a stress $\taur=22$Pa is first
applied during a time $t_{\scriptscriptstyle 0}=$22s just longer
than the liquid/solid transition time $\tt$=19s before being
lowered to zero. We observe that the elastic modulus value and its
evolution in time are exactly the same as when a stress
$\taur=22$Pa is applied during the whole experiment. This result,
together with the observation that the stress applied only while
the material is in a liquid state has basically no influence on
the elastic modulus value, means that the material state is
changed only by the stress applied during the very short moment
during which the material passes from a liquid to a solid state.
The strain of a few unities during which this liquid/solid
transition occurs is enough to induce the effect we observe.
Afterwards, when the material is in a solid state, this stress can
be removed with no consequence: the material has been irreversibly
changed. It also shows that the differences we observe in the
mechanical behavior as a function of the stress $\taur$ do not
simply reflect a non-linear elastic behavior: when $\taut=22$Pa,
the same elastic modulus is observed whether it is measured around
0Pa or around 22Pa during the aging
(Fig.~\ref{Fig_effet_cont_sur_elas2}).

\subsubsection*{\textbf{\textit{Impact of the stress applied during the aging
in the solid state}}}

Another important consequence of obtaining the same results with
the procedures of Fig.~\ref{fig_procedures}b and
Fig.~\ref{fig_procedures}d is that maintaining the stress $\taur$
during the whole aging after the liquid/solid transition does not
enhance the strengthening of the material. In other words, once
the material has been changed during the liquid/solid transition,
there is no stress-induced accelerated aging.

The material state thus seems to depend only on the stress applied
during the liquid/solid transition. However, another possibility
is that applying a shear stress at any time during the rest in the
solid state may lead to the same consequence: once the material is
in its solid regime, the stress may induce a reorganization of the
material structure that modifies its mechanical state once and for
all. This can be checked with the procedure of
Fig.~\ref{fig_procedures}e, where the liquid/solid transition
occurs under a zero stress whereas a stress $\taua\neq 0$ is
applied once the material is in its solid state. For the sake of
simplicity, we consider only stress histories in the solid regime
leading to a small creep flow, ensuring that the material stays in
a solid state. As the yield strain of the suspension is of order
0.1 (see Fig.~\ref{fig_yield_measurement}) a small creep flow is
defined here as a creep flow of strain less than 0.05. As the
static yield stress of the materials we study increases with the
resting time (see below), such simple histories can be achieved by
applying the aging stress $\taua$ only after a long enough resting
time (whose value depend on the stress and the material) at
$\taur=0$Pa. In Fig.~\ref{Fig_effet_cont_sur_elas2} we plot the
elastic modulus vs.\ time when first applying a stress $\taut=0$Pa
during 30s and then a stress $\taua=22$Pa during 300s (leading to
a creep flow of strain 0.03). With this procedure, we find that
the same modulus is measured when no stress is applied during both
the liquid/solid transition and the aging, as when a stress
$\taua\neq0$ is applied only once the material is in its solid
state. This finally shows that there is no stress-induced
accelerated mechanical aging in solid materials obtained with
$\taut=0$Pa.

\begin{figure}[htbp] \begin{center}
\includegraphics[width=7.9cm]{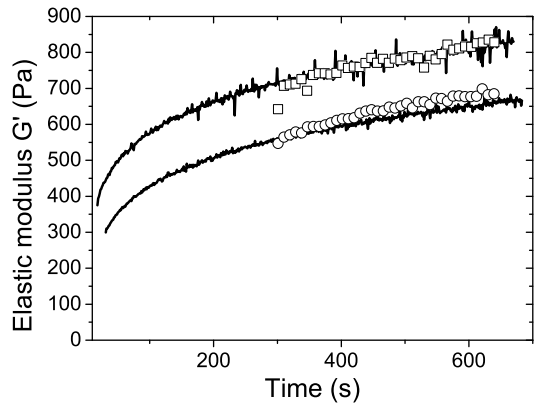}
\caption{Elastic modulus vs.\ time after strongly shearing a 6\%
bentonite suspension, measured with the various procedures of
Fig.~\ref{fig_procedures}: when a stress $\taur=0$Pa is applied
during the whole experiment (lower line); when a stress
$\taur=7$Pa is applied during the whole experiment (upper line);
when a liquid/solid transition stress $\taut=7$Pa is applied
during 30s and then lowered to zero during the aging (empty
squares); when a liquid/solid transition stress $\taut=0$Pa is
applied during 100s and then an aging stress $\taua=7$Pa is
applied during 200s (empty
circles).}\label{Fig_effet_cont_elas_4procedures}
\end{center} \end{figure}

We observe the same features in
Fig.~\ref{Fig_effet_cont_elas_4procedures} on a 6\% bentonite
suspension, with the procedures of
Fig.~\ref{fig_procedures}a,b,d,e. We first recover that the
elastic modulus is higher when $\taut\neq0$ than when $\taut=0$Pa.
We then recover that the same result is obtained when applying a
stress $\taur\neq0$ during both the liquid/solid transition and
the aging, and during the liquid/solid transition only. We also
recover that the material behavior is basically the same when no
stress is applied during both the liquid/solid transition and the
aging as when a stress $\taua\neq0$ is applied only once the
material is in a solid state. These results confirm that what
matters is actually the stress that is applied during the
liquid/solid transition: this stress only is at the origin of the
strengthening of the material in its solid state. The stress
applied during the aging has basically no influence on the solid
mechanical state of the material.

Note that stress histories applied after the liquid/transition
leading to a more important creep flow were found to lead to more
complex histories whose analysis is out of the scope of the
present paper. When such a stress history is applied to the
material, after the liquid/solid transition that occurs under a
naught stress, one first observes a solid/liquid transition
induced by the applied stress and then a second liquid/solid
transition. The stress that is applied during these phases may
then have an impact on the material state because it is applied
during a new liquid/solid transition.

\subsubsection*{\textbf{\textit{Impact of the stress history
on the static yield stress}}}

\begin{figure}[hbtp] \begin{center}
\includegraphics[width=7.9cm]{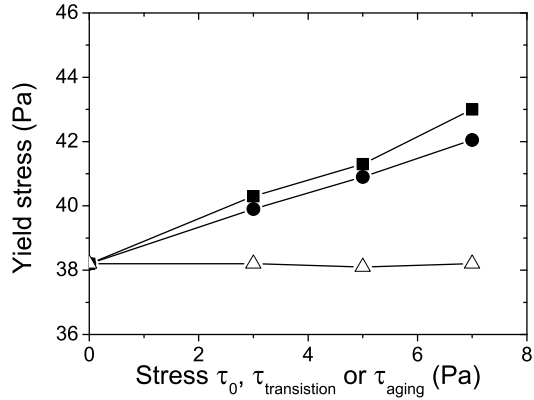}
\caption{Yield stress 300s after strongly shearing a 6\% bentonite
suspension, vs.\ $\taur$, $\taut$ or $\taua$, for the various
procedures of Fig.~\ref{fig_procedures}: when a stress $\taur$
ranging between 0 and 7Pa is applied (squares) during the whole
experiment; when a liquid/solid transition stress $\taut$ ranging
between 3 and 7Pa is applied during 30s and then removed
(circles); when a liquid/solid transition stress $\taut=0$Pa is
applied during 100s and then an aging stress $\taua$ ranging
between 3 and 7Pa is applied during 200s (open
triangles).}\label{Fig_effet_cont_sur_seuil_bento2}
\end{center} \end{figure}

We also observe that the static yield stress depends on the shear
stress history applied below the dynamic yield stress: if a
constant stress is applied after the preshear, the yield stress is
higher than when no stress is applied
(Fig.~\ref{Fig_effet_cont_sur_seuil_bento2}). The yield stress
measurements performed on a 6\% bentonite suspension with the
procedures of Fig.~\ref{fig_procedures}a,b,d,e are presented vs.\
the stress applied during the liquid/solid transition and/or the
aging in Fig.~\ref{Fig_effet_cont_sur_seuil_bento2}, for a total
aging time in the solid regime of 300s. As observed for the
elastic modulus measurements, we find that the same result is
obtained when applying a shear stress $\taur$ during both the
liquid/solid transition and the aging and during the liquid/solid
transition only. Moreover, when the aging stress $\taua$ is not
applied during the liquid/solid transition but only once the
material is in a solid state, then the yield stress is the same as
when no stress is applied during aging. This shows that, as for
the elastic modulus, the yield stress enhancement is induced only
by the stress applied during the liquid/solid transition. The
stress applied during the aging has basically no influence on the
solid mechanical state of the material.

\subsubsection*{\textbf{\textit{Summary}}}

To sum up, we have observed that the overall mechanical behavior
of a bentonite suspension depends strongly on the stress $\taut$
applied during its liquid/solid transition: the material is
strengthened when a non-zero $\taut$ is applied. On the other
hand, we observed basically no dependence on the shear history in
the liquid regime nor on the stress applied at rest. What matters
is thus applying a stress to the material during the few seconds
after the preshear during which a liquid/solid transition occurs,
until it is in a solid state. Then this stress can be removed with
no consequence: the material has been irreversibly changed. We
also found that the materials age, but, surprisingly, we found
that the mechanical aging kinetics (G' evolution in time) is
basically unchanged by the stress history.

\subsection{Constitutive laws accounting for the liquid/solid transition stress}

In the following, most results were obtained using the procedure
of Fig.~\ref{fig_procedures}d, i.e. a shear stress $\taut$ is
applied during the liquid/solid transition and is removed during
the aging of the material; the elastic modulus evolution in time
is then measured around a naught stress. Afterwards, the yield
stress is measured as a function of the time $\ta$ spent by the
materials in their solid regime. The elastic modulus measurements
are performed on all the materials presented in
Sec.\ref{section_materials}; the yield stress measurements are
performed only on the bentonite suspensions.

\subsubsection{Elastic modulus}

\begin{figure}[htbp] \begin{center}
\includegraphics[width=7.9cm]{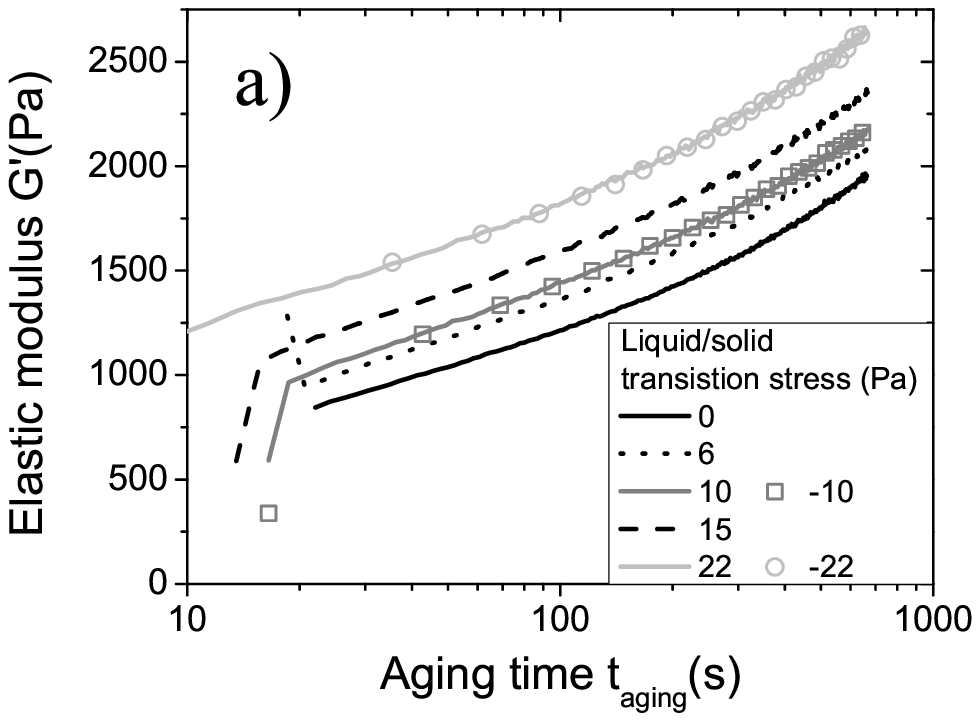}
\includegraphics[width=7.9cm]{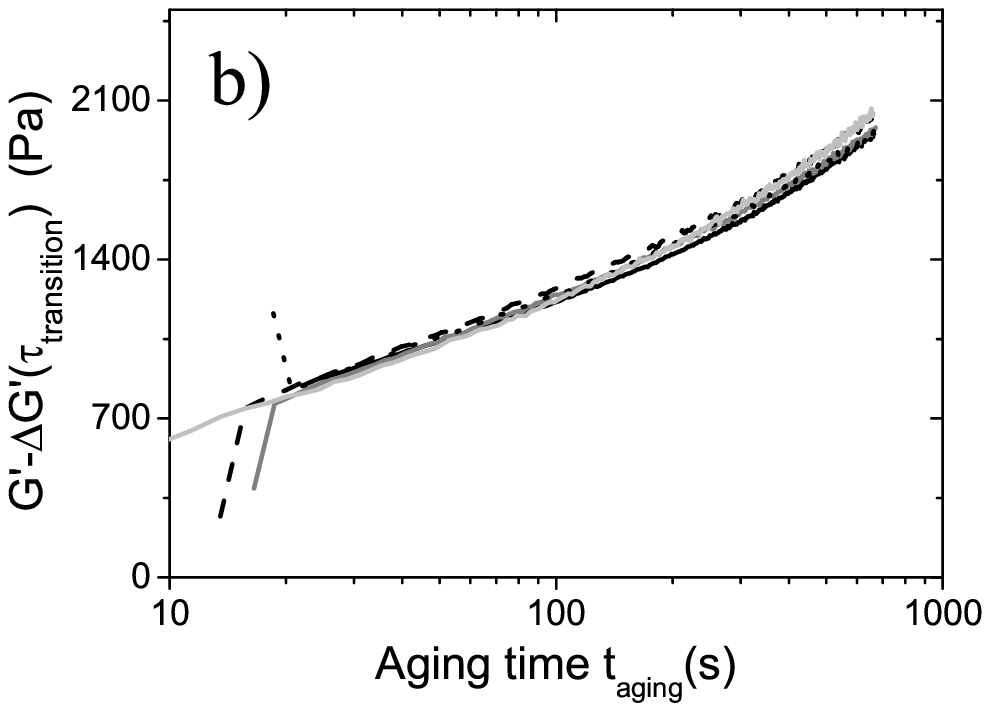}
\caption{a) Elastic modulus vs.\ time after strongly shearing a
9\% bentonite suspension, measured around 0Pa for various stresses
$\taut$ applied during the liquid/solid transition (a negative
value of $\taut$ represents a stress applied in the direction
opposite to the preshear direction). b) Same data as in
Fig.~\ref{Fig_effet_cont_sur_elas_bento1}a, when the elastic
modulus is shifted by a constant value $\Delta G'(\taut)$ (see
Fig.~\ref{Fig_bilan_delta_Gprime_ttes_bentos}).}\label{Fig_effet_cont_sur_elas_bento1}
\end{center} \end{figure}

\begin{figure}[htbp] \begin{center}
\includegraphics[width=7.9cm]{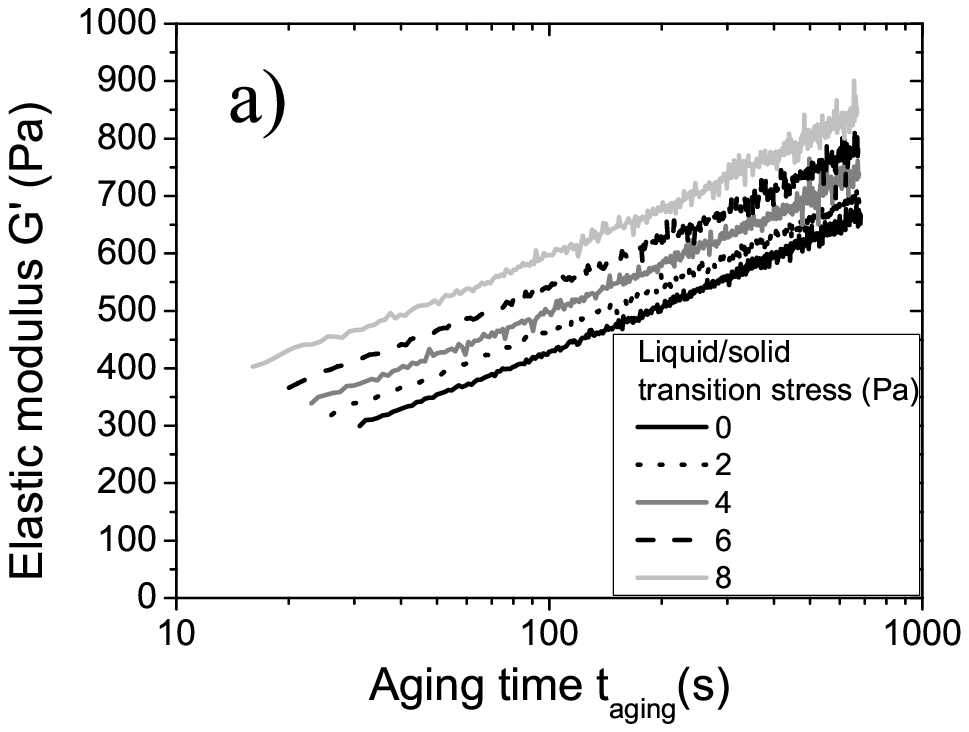}
\includegraphics[width=7.9cm]{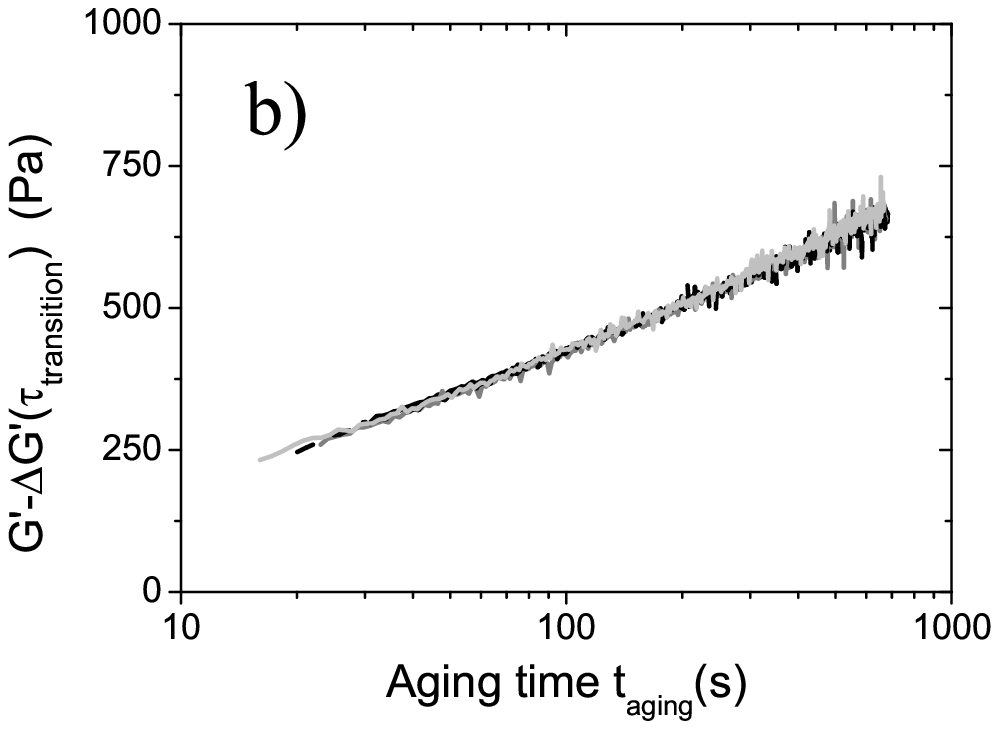}
\caption{Same plots as in
Fig.~\ref{Fig_effet_cont_sur_elas_bento1} for a 6\% bentonite
suspension.}\label{Fig_effet_cont_sur_elas_bento2}
\end{center} \end{figure}

\begin{figure}[htbp] \begin{center}
\includegraphics[width=7.9cm]{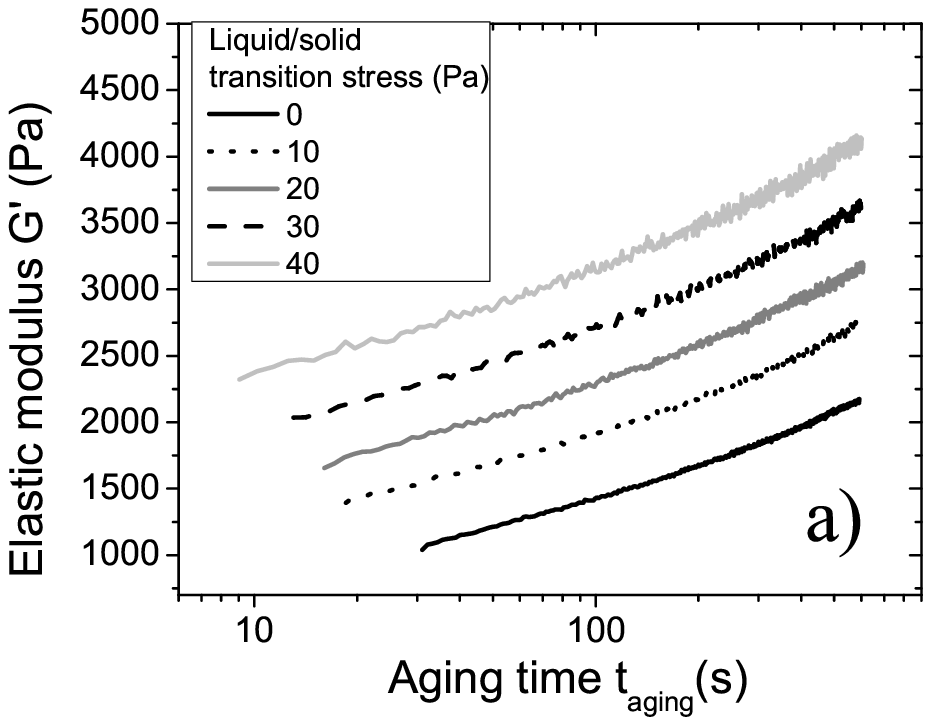}
\includegraphics[width=7.9cm]{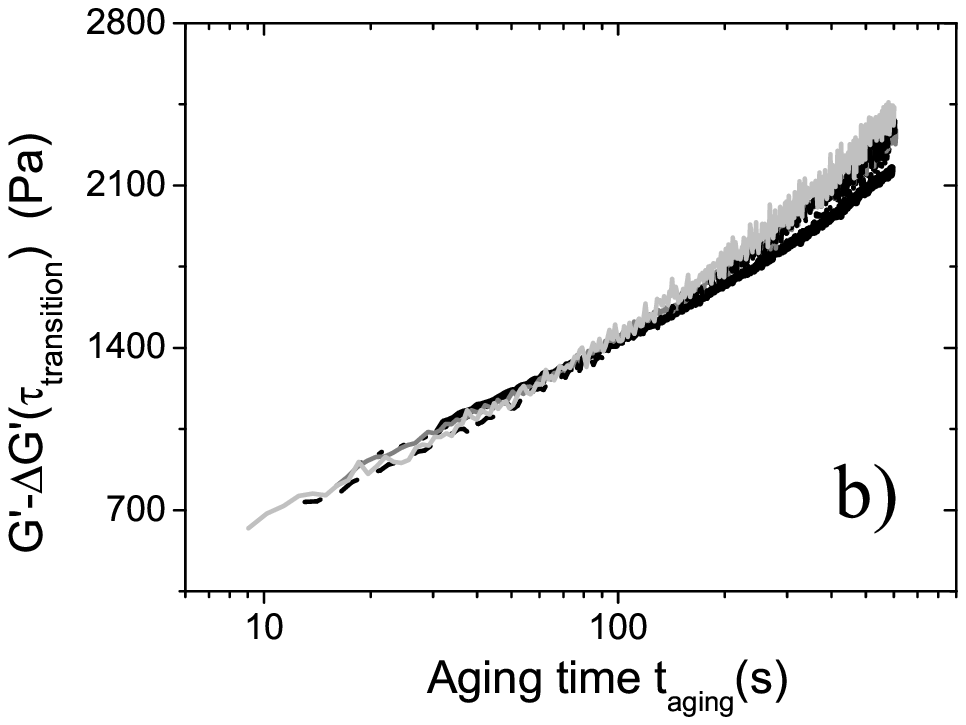}
\caption{Same plots as in
Fig.~\ref{Fig_effet_cont_sur_elas_bento1} for a 10\% bentonite
suspension.}\label{Fig_effet_cont_sur_elas_bento3}
\end{center} \end{figure}

In Fig.~\ref{Fig_effet_cont_sur_elas_bento1}a,
\ref{Fig_effet_cont_sur_elas_bento2}a, and
\ref{Fig_effet_cont_sur_elas_bento3}a, we plot the elastic modulus
values vs.\ the time spent in the solid regime for different
values of $\taut$ applied during the liquid/solid transition, for
3 different bentonite suspensions.

For all materials, we notice that the elastic modulus is strongly
increased by the stress that is applied during the liquid solid
transition. E.g., on Fig.~\ref{Fig_effet_cont_sur_elas_bento3}a,
we see that the elastic modulus measured after a 100s aging time
increases from 1400 to 3000Pa when $\taut$ is changed from 0 to
40Pa. However, surprisingly, as noticed in
Sec.~\ref{section_elasticity}, we observe that the aging kinetics
seems to be basically unchanged by this stress. The elastic
modulus value $G'(\ta,\taut)$ as a function of the time $\ta$
spent in the solid regime and of the stress $\taut$ applied during
the liquid/solid transition would then read \bea
G'(\ta,\taut)=G'_{\scriptscriptstyle 0}(\ta)+\Delta
G'(\taut)\label{eq_scaling_gprime} \eea

\begin{figure}[htbp] \begin{center}
\includegraphics[width=7.9cm]{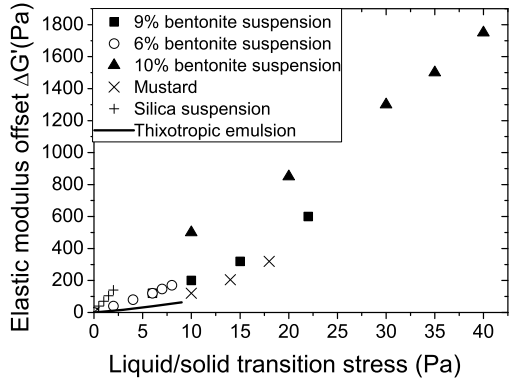}
\caption{Elastic modulus offset $\Delta G'(\taut)$ obtained by
superposing the data of Fig.~\ref{Fig_effet_cont_sur_elas_bento1}
(squares), \ref{Fig_effet_cont_sur_elas_bento2} (open circles),
\ref{Fig_effet_cont_sur_elas_bento3} (triangles), and
\ref{Fig_effet_cont_sur_elas_other_materials}a (crosses) and
\ref{Fig_effet_cont_sur_elas_other_materials}b (plus) vs.\ the
stress $\taut$ applied during the liquid/solid
transition.}\label{Fig_bilan_delta_Gprime_ttes_bentos}
\end{center} \end{figure}

In Fig.~\ref{Fig_effet_cont_sur_elas_bento1}b,
\ref{Fig_effet_cont_sur_elas_bento2}b, and
\ref{Fig_effet_cont_sur_elas_bento3}b, we now plot the elastic
modulus values shifted by constant values $\Delta G'(\taut)$ vs.\
the time spent in the solid regime for the same data as in
Fig.~\ref{Fig_effet_cont_sur_elas_bento1}a,
\ref{Fig_effet_cont_sur_elas_bento2}a, and
\ref{Fig_effet_cont_sur_elas_bento3}a. We observe a rather good
superposition of all data; there may be, however, a small
discrepancy beyond a 15min aging time as seen on
Fig.~\ref{Fig_effet_cont_sur_elas_bento3}b, but at this stage this
is a second-order effect. The $\Delta G'(\taut)$ values for all
$\taut$ values on all materials are plotted in
Fig.~\ref{Fig_bilan_delta_Gprime_ttes_bentos}.

We observe that, for a given material, $\Delta G'(\taut)$
increases linearly with $\taut$. However, writing \bea\Delta
G'(\taut)=\alpha\taut\label{eq_scaling_deltagprime}\eea we see
that there is no universal value of $\alpha$. $\alpha$ seems to
depend on the clay particles volume fraction: we find $\alpha$
values between 20 and 40. It is still possible that the relevant
scaling is $\Delta G'(\taut)/G'_m\approx\lambda\,\taut/\tau_m$,
with $\lambda$ a universal constant, and with $G'_m$ and $\tau_m$
mechanical characteristics of the materials: $\tau_m$ may then be
the dynamic yield stress, but, as the materials age, we did not
find how to define a characteristic elastic modulus $G'_m$,
therefore we could not test such a scaling.

\begin{figure}[htbp] \begin{center}
\includegraphics[width=7.9cm]{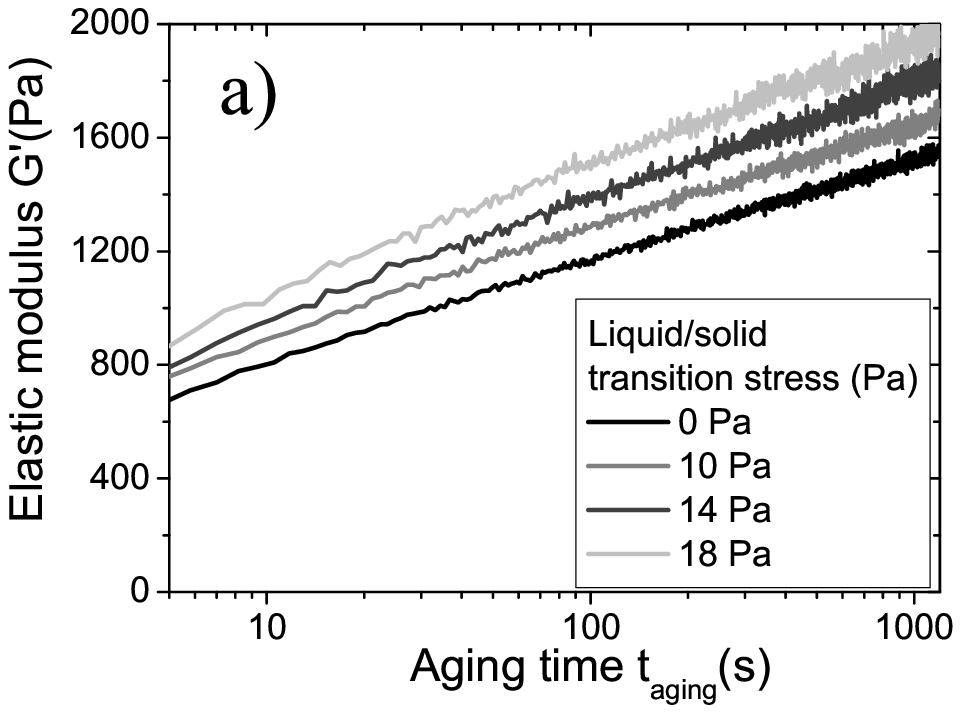}
\includegraphics[width=7.9cm]{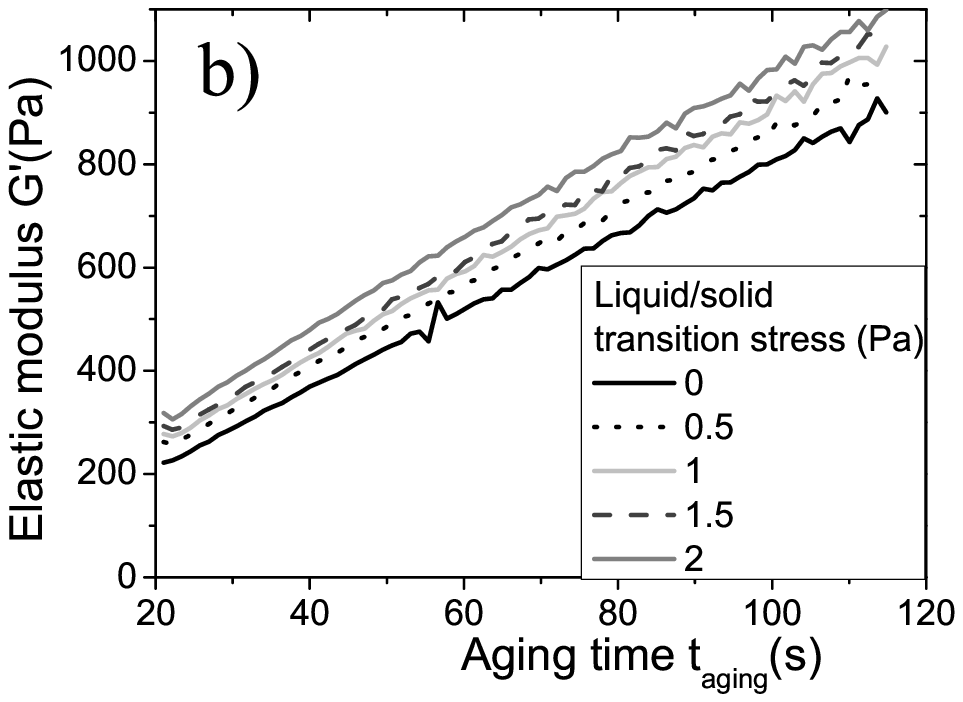}
\includegraphics[width=7.9cm]{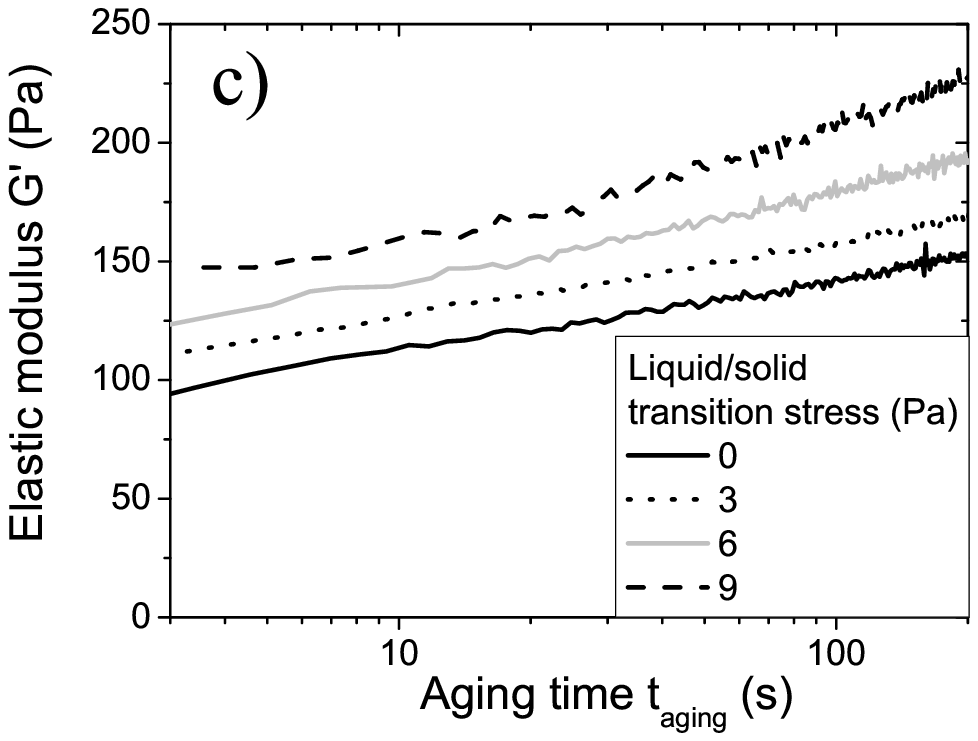}
\caption{a) Elastic modulus vs.\ time for various stresses
$\taur<\tau_d$ applied after strongly shearing a mustard. b)
Elastic modulus vs.\ time for various stresses $\taur<\tau_d$
applied after strongly shearing a silica suspension. c) Elastic
modulus vs.\ time for various stresses $\taur<\tau_d$ applied
after strongly shearing a thixotropic
emulsion.}\label{Fig_effet_cont_sur_elas_other_materials}
\end{center} \end{figure}

In order to show that the phenomenon we observe here is not
specific to bentonite suspensions, but is a general phenomenon
occurring in a wide range of thixotropic materials, we performed
the same experiments (although in a less detailed way) on a
mustard, a silica suspension, and a thixotropic emulsion (see
Sec.~\ref{section_materials}). The results of the procedures of
Fig.~\ref{fig_procedures}a and \ref{fig_procedures}b are depicted
in Fig.~\ref{Fig_effet_cont_sur_elas_other_materials}.

We observe that, as in the bentonite suspensions, the elastic
modulus values of the mustard, the silica suspension, and the
thixotropic emulsion strongly increase when a stress $\taut$ is
applied during the liquid/solid transition. We checked that the
same result is recovered when the stress is applied only during
the liquid/solid transition as when it is applied during both the
liquid/solid transition and the aging. We also observe in
Fig.~\ref{Fig_effet_cont_sur_elas_other_materials} that the aging
kinetics is basically unmodified by the applied stress (there may
be a very small increase of the elastic modulus increase rate of
the silica suspension and the thixotropic emulsion for high
stresses, but we did not study this effect and it is clearly a
second-order effect). The effect of $\taut$ on the quantitative
increase of the elastic modulus of these materials is depicted in
Fig.~\ref{Fig_bilan_delta_Gprime_ttes_bentos}: it is of the same
order as in the bentonite suspensions, and it is consistent with
Eq.~\ref{eq_scaling_deltagprime}. However, the $\alpha$ parameter
of Eq.~\ref{eq_scaling_deltagprime} is of the order of 80 for the
silica suspension whereas it is of order 16 for the mustard, and
of order 7 for the thixotropic emulsion showing again that this
parameter is material dependent.

Finally, we performed the same experiments on a simple
(non-thixotropic) emulsion. In contrast with what is observed in
the loaded emulsion, we found that elastic modulus of the simple
emulsion has the same value whatever the stress $\taut$. It shows
that the effect we observe is certainly specific to thixotropic
materials, and it points out the role of the dynamics of links
creation between the particles.

\subsubsection{Yield stress}\label{section_yield_stress}

\begin{figure}[htbp] \begin{center}
\includegraphics[width=7.9cm]{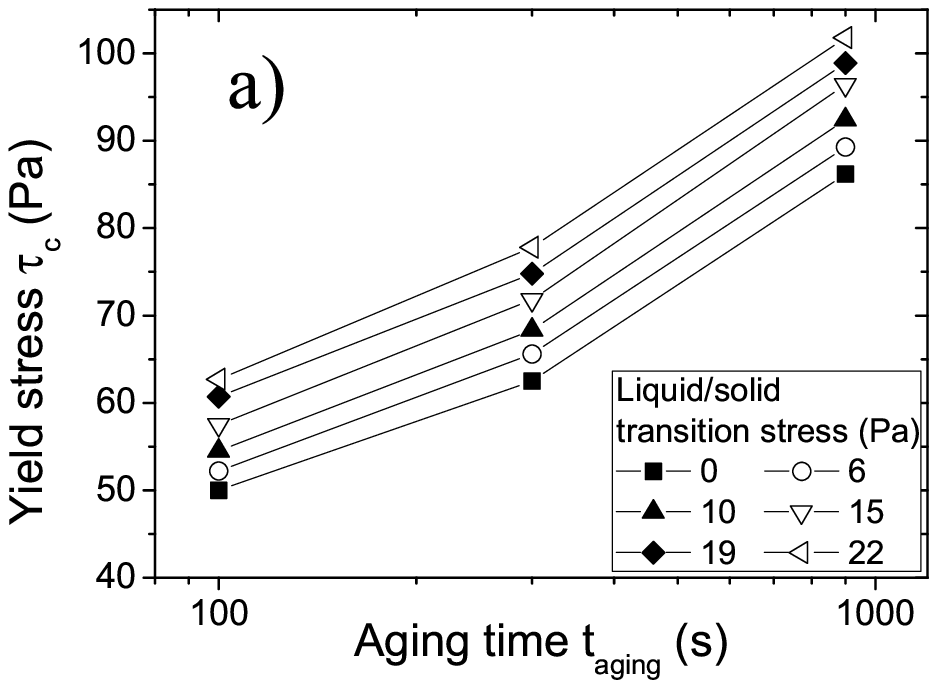}
\includegraphics[width=7.9cm]{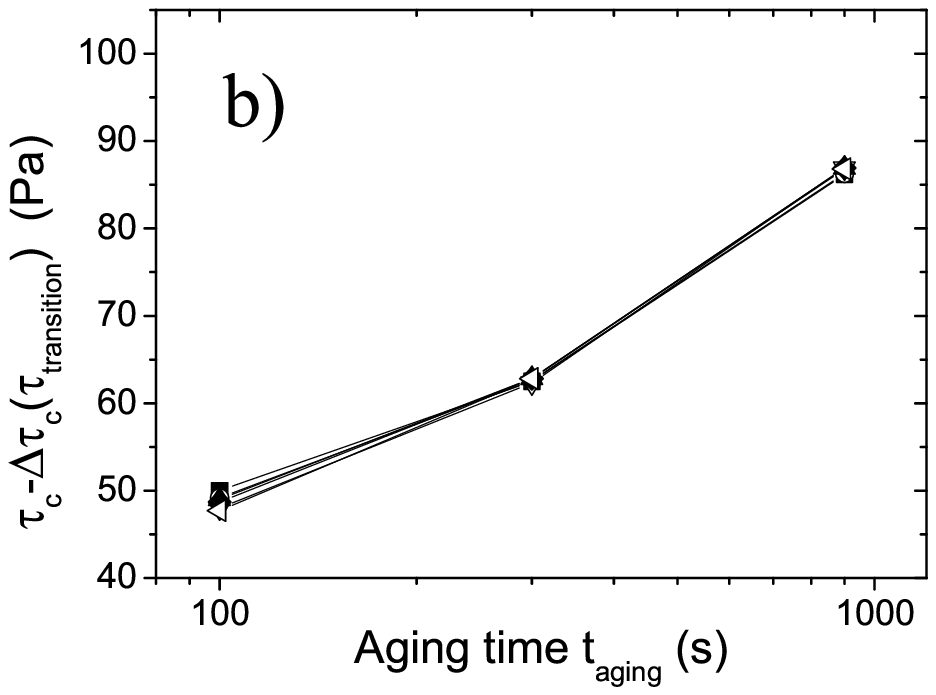}
\caption{a) Yield stress $\tau_c$ vs.\ time after strongly
shearing a 9\% bentonite suspension, for various stresses $\taut$
applied during the liquid/solid transition. b) Same data as in
Fig.~\ref{Fig_effet_cont_sur_seuil_bento1}a, when the yield stress
is shifted by a constant value $\Delta
\tau_c(\taut)$.}\label{Fig_effet_cont_sur_seuil_bento1}
\end{center} \end{figure}

In Fig.~\ref{Fig_effet_cont_sur_seuil_bento1}, we present the
yield stress values measured on a 9\% bentonite suspension for
various stresses $\taut$ ranging between 0 and 22Pa, and for
various aging times.

In Fig.~\ref{Fig_effet_cont_sur_seuil_bento1}, we observe that the
yield stress is strongly increased by the stress $\taut$ that is
applied during the liquid/solid transition. However, as already
noticed in Sec.~\ref{section_elasticity} for the elastic modulus
measurements, we observe that the aging kinetics seems to be
unchanged by this stress. The yield stress value
$\tau_c(\ta,\taut)$, as a function of the time $\ta$ spent in the
solid regime and of the stress $\taut$ applied during the
liquid/solid transition, would then just be shifted by a constant
$\Delta\tau_c(\taut)$ value and read \bea
\tau_c(\ta,\taut)=\tau_{c_{\scriptscriptstyle 0}}(\ta)+\Delta
\tau_c(\taut)\label{eq_scaling_yield_stress} \eea

\begin{figure}[htbp] \begin{center}
\includegraphics[width=7.9cm]{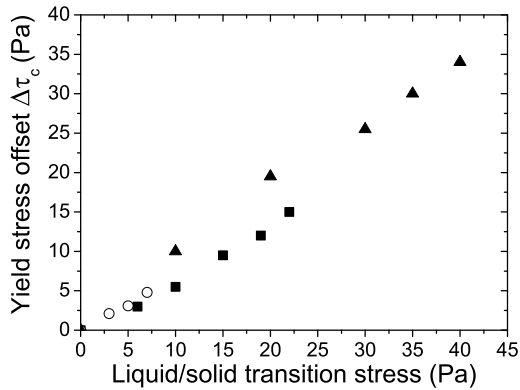}
\caption{Yield stress offset $\Delta\tau_c(\taut)$ obtained by
superposing the data of Fig.~\ref{Fig_effet_cont_sur_seuil_bento1}
for a 9\% bentonite suspension (squares), and for the same
experiments on a 6\% (open circles) and a 10\% (triangles)
bentonite suspension, vs.\ the stress $\taut$ applied during the
liquid/solid transition.}\label{Fig_bilan_delta_seuil_ttes_bentos}
\end{center} \end{figure}

We performed the same experiments on the two other bentonite
suspensions and observed the same features: all results are in
agreement with Eq.~(\ref{eq_scaling_yield_stress}). The stress
offsets $\Delta \tau_c(\taut)$ for all bentonite suspensions are
presented in Fig.~\ref{Fig_bilan_delta_seuil_ttes_bentos}.

We observe in Fig.~\ref{Fig_bilan_delta_seuil_ttes_bentos} that
the stress shift $\Delta \tau_c(\taut)$ increases linearly with
$\taut$. However, writing
\bea\Delta\tau_c(\taut)=\beta\taut\label{eq_scaling_deltayield_stress}
\eea we see, as for the elastic modulus measurements, that there
is no universal value of $\beta$. $\beta$ seems to depend slightly
on the material: we find $\beta$ values between 0.6 and 0.9.

It is worth noting that the phenomenon we observe is not an
artefact of the yield stress measurement procedure: we checked
that the same phenomenon, with the same quantitative effect, is
observed with another yield stress measurement method, based on a
shear stress ramp \cite{uhlherr2005,Ovarlez2007b}.

Finally, note that measuring the yield stress by imposing a slow
velocity in the opposite direction to the preshear yields a
slightly higher value than measuring this yield stress in the same
direction as the preshear. This is due to a shear-induced
anisotropy of the suspensions \cite{parsi1987,mathis1988}.
However, we checked that the effect of $\taut$ is to increase the
yield stress by the same amount whatever the relative directions
of the preshear and the yield stress measurement. This indicates
that the yield surface is increased isotropically rather than
shifted.

\section{Discussion}\label{section_discussion}

\subsection{Interpretation}

In order to account for all the observations we made in
Sec.~\ref{section_mechanics}, we need to find a structuration
mechanism that (i) depends on the shear stress $\taut$ applied at
the transition between the liquid and the solid regime so that
different initial solid mechanical states are created by different
$\taut$, (ii) is consistent with the mechanical properties being
shifted by a value roughly proportional to $\taut$
(Eqs.~(\ref{eq_scaling_deltagprime}) and
(\ref{eq_scaling_deltayield_stress})), (iii) is consistent with a
structuration rate (the increase rate of their elastic modulus and
yield stress) basically independent of the stress history, (iv) is
consistent with the different microstructures of the various
suspensions we study.

Point (i) can be easily understood. It is well known that a flow
induces an anisotropic microstructure in suspensions of colloidal
and noncolloidal particles \cite{mathis1988,parsi1987}, and that
the degree of anisotropy increases with the shear rate as a result
of a competition between Brownian motion (that tends to isotropize
the microstructure) and hydrodynamic interactions \cite{Foss2000}.
Here, we face complex suspensions that jam rapidly when the
applied stress is lower than the dynamic yield stress $\tau_d$. We
can plausibly propose that if this structuration occurs while a
stress is applied to the material, different microstructures are
frozen during the liquid/solid transition depending on $\taut$. As
the mechanical properties of suspensions depend on their
microstructure \cite{Russel-Saville-Schowalter-1995}, these
materials then have naturally different elastic moduli. However,
it is important to note that the solid mechanical properties of
the materials depend mainly on the stress applied at the
liquid/solid transition, while there is basically no influence of
the shear history in the liquid regime. This would mean that even
if the flow induces a microstructure anisotropy, this anisotropy
is likely to disappear quickly if the stress is removed before the
liquid/solid transition, due to Brownian motion of the particles
in the liquid state. A key point thus seems that this anisotropy
is frozen at the liquid/solid transition thanks to the stress that
is applied during this transition. Actually, the particle
positions in the disordered particle network that is frozen must
be compatible with the stress applied during the transition. This
means that, depending on this stress, preferential orientations of
the pair forces are chosen to ensure this compatibility. In
Appendix~\ref{section_appendix}, leaving aging apart, we present a
sketch of a simple 2D model colloidal suspension in order to
present how such a mechanism may work. Through a micromechanical
analysis of the behavior of this model suspension, we simply draw
the consequences of the existence of repulsive and attractive
forces in the suspension. Then, within this picture, we show that
a jamming at a given $\taut$ may induce a change in the
microstructure that actually implies a roughly linear dependence
of the effective elastic modulus on $\taut$.

\subsubsection*{\it The role of aging}

As regards the mechanism of aging, it should first be noted that
the observation of similar evolution in time of the moduli of
materials for different $\taut$ would mean that although these
materials behave differently on the mechanical point of view, the
underlying physical mechanism of their aging is the same and is at
the same stage at a same given aging time $\ta$. Then, the
requirement (iii) that the increase rate of the mechanical
properties is basically independent of $\taua$ is important and
discriminating. It seems that the energy landscape visited during
the aging is unaffected by the shear stress applied during aging
in our systems, and that a model of aging based on stress-biased
energy barriers \cite{Berthoud1999} would fail to describe our
systems. This is thus probably neither consistent with models
based on the aging at the contact scale \cite{manley2005} nor with
models in which the yield stress originates from the static
friction between the particles \cite{Furst2007} as it was observed
in several systems that the aging of solid contacts is very
sensitive to the shear stress applied during aging
\cite{Berthoud1999,Losert2000,Restagno2002,Ovarlez2003}. This
feature actually requires an aging mechanism that is basically
insensitive to the stress transmitted in the solid network. We
think that this is consistent with a structure build-up due to
creation of new contacts between particles, leading to
strengthening of the material. In this case, once the material is
in a solid state, we can think that the stress $\taua$ is
transmitted only by the solid skeleton: the free particles (or
aggregates) in the suspension that stick on this skeleton thanks
to Brownian motion are then insensitive to the applied stress. As
a consequence, the further increase of the elastic modulus due to
the new links that are created would be independent of the stress
$\taua$ applied during the aging.

\subsection{Macroscopic consequences}\label{section_macroscopic}

From a practical point of view, our results point out the
importance of well controlled experimental conditions for
performing yield stress measurements. On the one hand, the yield
stress at which a flow stops (the dynamic yield stress $\tau_d$)
is well defined from creep tests \cite{Coussot2006}: starting from
a fully destructured liquid state of a material, if a creep stress
above $\tau_d$ is prescribed then the material flows steadily
whereas for a creep stress below $\tau_d$ there is a creep flow
that is slowing down at any time and the strain rate tends towards
0. On the other hand, we have observed a new feature: the yield
stress at which the flow starts (the static yield stress $\tau_c$)
is not uniquely defined: it depends strongly on the stress applied
during the liquid/solid transition.

As a consequence, if one just pours a material in a cup to perform
a measurement, how it was poured, i.e. how its flow stopped, may
have an influence on the yield stress that is measured.
Nevertheless, in order to avoid irreproducible material
preparation and to perform measurements on a well defined state,
one usually preshear the material, and then stops shearing the
material during a given resting period before performing the yield
stress measurement. However, an important difference remains
between different measurement methods. This difference stands in
what is called 'rest': a material is said to be at rest when it is
not flowing. With a stress-controlled rheometer, the rest is
imposed by applying a stress $\taur=0$Pa; as we have shown in this
paper, this yields a well defined unique solid state. With a
rate-controlled rheometer, after a preshear at a given $\gdot_p$,
the rest is defined by $\gdot_{rest}=0$s$^{-1}$ (i.e. the tool
rotation is abruptly stopped) so that the stress state during the
liquid/solid transition and rest is not {\it a priori} known. As
we have shown that the static yield stress actually depends on
this stress, this implies that the initial state of the material
is ill-defined in such experiments. This ill-definition of the
initial state can be seen in papers showing raw data of yield
stress measurements performed with the vane method (see e.g.
Fig.~4 of \cite{dzuy1985} and Fig.~9 of \cite{james1987}): in this
case, it is observed that the stress at the beginning of the
measurement is different from zero, which means that the material
was stressed during the flow stoppage and the rest. This feature
may explain partly the differences between measurements performed
with stress controlled rheometers {\it and} strain controlled
rheometers, and between different groups in \cite{nguyen2006} on
the same materials. This shows that stress-controlled rheometers,
or rate and stress-controlled rheometers are preferable for well
defined yield stress measurements: an appropriate resting period
should be defined as a period during which a stress $\taur=0$Pa is
applied rather than a shear rate $\gdot_{rest}=0$s$^{-1}$.

While our paper was under review, we read a preprint
\cite{osuji2007} that shows results that may be consistent with
ours on a colloidal gel. In this preprint, \citet{osuji2007} found
that when the flow of their system is stopped at $\gdot=0$s$^{-1}$
after a strong preshear at $10^2$-$10^3$s$^{-1}$, the elastic
modulus $G'$ at rest is higher for a higher intensity of preshear.
They attributed this increase to the dependence of the fractal
structure on the shear stress $\tau_{preshear}$ applied during the
preshear. However, they also noticed that the elastic modulus is
linearly correlated with what they call 'residual' or 'internal'
stress $\tau_{internal}$, namely the stress resulting from the
rapid quench in the solid regime when applying abruptly
$\gdot=0$s$^{-1}$ at the end of the preshear. We think that
$\tau_{internal}$ is nothing else than our liquid/solid transition
stress $\taut$ which is here uncontrolled because of the use of a
velocity-controlled mode to stop the shear. We suggest that the
relationship found by \citet{osuji2007} between $G'$ and
$\tau_{preshear}$ may be fortuitous and simply due to the fact
that for a given $\tau_{preshear}$, they have a given uncontrolled
$\taut$ (or equivalently $\tau_{internal}$) upon imposing
$\gdot=0$s$^{-1}$. We also suggest that the relationship between
$G'$ and $\taut$ (or $\tau_{internal}$) is more relevant. This can
be easily proved by varying independently $\tau_{preshear}$ and
$\taut$, i.e. by working with a stress controlled mode, at least
for flow stoppage.

This phenomenon may also explain the discrepancy between classical
rheometrical measurements and inclined plane measurements with the
method of \citet{Coussot2002}, performed on the same materials by
\citet{nguyen2006} (see Fig.~9 of \cite{nguyen2006}).
\citet{Coussot2002} proposed to first pour the material on an
inclined plane at a given slope in order to measure its dynamic
yield stress (from the final height of the deposit), and then to
incline further the plane in order to measure its static yield
stress (from the angle of the plane at flow restart). Our
experiments show that this method is incorrect: when the flow
stops on the inclined plane, the stress at the base of the
material is equal to its dynamic yield stress: the liquid/solid
transition then occurs under a non-zero shear stress. The static
yield stress of the material measured afterwards is then higher
than what it would be if the liquid/solid transition had occurred
under a zero shear stress. Moreover, as the stress increases from
bottom to top in the layer, the solid material is heterogeneous:
its actual yield stress increases from bottom to top. We will
present a detailed study of this problem, as well as a relevant
method to measure correctly the yield stress in an inclined plane
experiment in a future work \cite{Ovarlez2007b}.

More generally, the way the flow is stopped before a yield stress
measurement should depend on the practical application that is of
interest: in problems where one seeks for the restart of a flow
after an abrupt flow cessation, it is of importance to know how
the previous flow actually stopped. E.g., in pipe flows such as
concrete pumping or in extrusion processes, which are often rate
controlled through the action of a piston, the stress at a flow
cessation is likely to be near the dynamic yield stress: the flow
will then be harder to restart than what would be expected from a
rheometrical measurement performed with $\taut=0$Pa. On the other
hand, in applications where the material passes from a liquid to a
solid state under a naught shear stress then the relevant yield
stress is the one measured with $\taut=0$Pa. This is the case e.g.
when a material is poured in its liquid state in a receptacle: in
this case, the liquid material is in a hydrostatic stress state
when it passes from a liquid to a solid state. An example of such
situation is given by thixotropic concrete casting in formwork: in
order to predict the stresses supported by the walls after
concrete casting, one needs the static yield stress of the
material \cite{Ovarlez2006}, which is given in this case by the
one measured with $\taut=0$Pa; we actually showed that the yield
stress of thixotropic cement pastes involved in concrete mix
design strongly depends on $\taut$ \cite{Ovarlez2007c}.

Finally, it should be stated that for complex flow histories, a
given material stops under a heterogeneous stress distribution:
this yields to a material of heterogeneous mechanical properties.
Any flow restart may then happen with a complex yield surface
\cite{Ovarlez2007b}.

\section{Conclusion}
We have studied the mechanical aging at rest of several different
thixotropic colloidal suspensions, under various shear stress
histories applied during their flow stoppage and their aging in
their solid state. We have shown that their solid mechanical
properties depend strongly on the shear stress applied while they
pass from a liquid to a solid state (i.e. during flow stoppage).
Basically, we found that the elastic modulus and the yield stress
increase linearly with the shear stress applied at the
liquid/solid transition. On the other hand, we have shown that
there is negligible dependence of these mechanical properties on
the preshear history and on the shear stress applied at rest.
Compared with the impact of the shear stress applied at the
liquid/solid transition, applying the same stress only before or
only after the liquid/transition may induce only second-order
effects on the solid mechanical properties. Moreover, the elastic
modulus and the yield stress of thixotropic suspensions at rest
increase in time, but we have observed that the structuration rate
(the increase rate of the elastic modulus and the yield stress)
hardly depends on the stress history. We showed that the new
phenomenon we evidence in this paper may reflect the differences
in the microstructures that are frozen at the liquid/solid
transition. A micromechanical analysis of the behavior of a 2D
model colloidal suspension, with repulsive and attractive pair
interactions, allowed us to show that a jamming at a given $\taut$
may induce a change in the microstructure that implies a roughly
linear dependence of the effective elastic modulus on $\taut$.
This points out the role of the internal forces in the colloidal
suspensions behavior. The independence of the increase rate of the
mechanical properties on the stress applied during aging may
require an aging mechanism that is insensitive to the stress
transmitted in the solid network: this is consistent with a
structure build-up through new contacts creation. We have also
shown that these results may have important macroscopic
consequences. This tells us how careful rheometrical test must be
designed to measure correctly the static yield stress; in
particular, one would better control the shear stress applied at
flow stoppage. Moreover, in any practical case, this implies that
the threshold for flow initiation as well as the yield surface in
the material depend a lot on how the flow was stopped.

\acknowledgments We thank Philippe Coussot, Ana\"el Lema\^itre,
and Nicolas Roussel for many fruitful discussions throughout this
work. Special thanks to Alexandre Ragouilliaux for providing the
thixotropic and simple emulsions. Support from the Agence
Nationale de la Recherche (ANR) is acknowledged (grant
ANR-05-JCJC-0214).

\appendix
\section{Impact of the liquid/solid transition stress in a simple sketch of a colloidal suspension}\label{section_appendix}

In the following, we present a simple model of a colloidal
suspension, in order to show how a liquid/solid transition
occurring at a given $\taut$ may induce a change in the
microstructure that implies a roughly linear dependence of the
effective elastic modulus on $\taut$. Of course, this model is not
intended to be a model of the systems we study experimentally;
moreover, note that we do not attempt to describe the evolution in
time of the microstructure in the solid regime. We intentionally
stay at a conceptual level as our aim is not to explain and
account for the time dependent properties of materials. It is just
a simple colloidal model system that allows us to draw the
consequences of the existence of interaction forces at a
macroscopic level, and to show that a simple micromechanical
analysis is able to account for the effect we observe
(independently of any aging feature). Of course, in order to model
our experimental systems, we would have to add complexity in the
model in order to take into account spatial heterogeneities,
aging, etc... But these last features, as we show in the
following, are not necessary to explain the observation of the
phenomenon we evidence: the key point is the existence of internal
forces in the suspension.

Our model colloidal suspension is a bidimensional monodisperse
suspension of particles distributed in an incompressible Newtonian
fluid. The particles interact through pair colloidal interactions
involving both attractive and repulsive forces
\cite{Russel-Saville-Schowalter-1995}. We focus on the mechanical
behavior at the liquid/solid transition: in this case, the
suspension is at rest and the hydrodynamic interactions are
negligible. We consider cases where the interaction forces
dominate the Brownian effect. Then, the particles form a
disordered network and the contribution of the fluctuations to the
stress tensor can be neglected.

For the sake of simplicity, it is assumed (i) that all the
repulsive forces (respectively all the attractive forces) have
same intensity equal to $\fp$ (resp. to $\fm > 0$), (ii) that the
distance between two interacting particles is equal to $\lp$
(resp. $\lm$) when the force is attractive (resp. repulsive).

Let us consider a representative elementary volume $V$ of the
suspension, large enough to be of typical composition. If the 2D
colloidal network is at rest, the macroscopic Cauchy stress tensor
reads~\cite{Batchelor-1970,Chateau-2001i,Russel-Saville-Schowalter-1995}:

\begin{equation}
  \label{eq:cauchy}
  \begin{array}{lll}
  \sigma_{xx} & = & {\displaystyle \frac{1}{|V|} \int_{-\pi/2}^{\pi/2}
    \left(\fp \lp \pp - \fm \lm \pm \right) \cos^2 \theta \, d \theta
  }\\
\\
  \sigma_{yy} & = & {\displaystyle \frac{1}{|V|} \int_{-\pi/2}^{\pi/2}
    \left(\fp \lp \pp - \fm \lm \pm \right) \sin^2 \theta \, d \theta
  }\\
\\
 \sigma_{xy} & = & {\displaystyle \frac{1}{|V|} \int_{-\pi/2}^{\pi/2}
   \left(\fp \lp \pp - \fm \lm \pm \right) \cos \theta \sin  \theta \,
   d \theta  }
  \end{array}
\end{equation}
where $\theta$ denotes the angular position of the vector joining
two interacting particles with respect to the $x$ axis, $|V|$ is
the area of the domain $V$ and $\pp d \theta$ (resp. $\pm d
\theta$) is the number of attractive (resp. repulsive) doublets
located in the representative elementary volume, of orientation
belonging to $[\theta, \theta + d \theta]$.

The overall behavior of the suspension is determined in the
framework of a mean field approach. Then, when a small macroscopic
shear strain $\gamma$ is prescribed to the representative
elementary volume, the variation of the distance between two
particles interacting through an attractive potential reads: $
\frac {d \lp}{\lp} = \gamma \cos \theta \sin \theta$. Of course,
the same relation holds for a repulsive doublet.

If $\kp$ (resp. $\km$) denotes the rigidity associated to the
attractive (resp. repulsive) potential when the distance between
the particles is equal to $\lp$ (resp. $\lm$), then the variation
of the elastic energy stored in the representative elementary
volume associated with the overall shear deformation is the sum of
the variation of elastic energy stored in all the doublets, which
reads:
\begin{equation}
  \label{eq:dW}
  d W = {\displaystyle \frac{1}{|V|} \int_{-\pi/2}^{\pi/2}
    \left[
        \left( \fp d \lp + \frac{1}{2} \kp d\lp^2 \right) \pp
    +   \left( \fm d \lm + \frac{1}{2} \km d\lm^2 \right) \pm
  \right]  \, d \theta
  }
\end{equation}
The overall behavior of the suspension being elastic,
Eq.~(\ref{eq:dW}) also reads $d W = \tau \gamma + \frac{1}{2} G
\gamma^2$, where $\tau$ denotes the overall shear stress (below
the yield stress) applied to the unstrained suspension and $G$ the
macroscopic shear modulus in the $x y$ direction.

This finally yields the following expression for the macroscopic
shear modulus:
\begin{equation}
  \label{eq:Ghom}
  G = \frac{1}{|V|}   \int_{-\pi/2}^{\pi/2}
  \left[\kp\,\lp^2\,\pp  + \km\,\lm^2\,\pm \right] \cos^2
    \theta \sin^2 \theta \; d\theta
\end{equation}

Let us now show that in this simple model colloidal suspension,
the behavior observed in our experiments is recovered. First, let
$\ppo$ and $\pmo$ denote the angular repartition of the doublets
in the solid state when no stress is applied during the
liquid-solid transition. The two functions $\ppo$ and $\pmo$ have
to ensure that the three integrals of Eq.~(\ref{eq:cauchy}) are
naught. The value of the shear modulus $G_0$ corresponding to this
case is obtained by putting $\ppo$ and $\pmo$ into
Eq.~(\ref{eq:Ghom}).

Many solutions for the orientation distribution of the doublets
exist so that Eqs.~(\ref{eq:cauchy}) are satisfied. We do not need
at this stage to make any assumption on this distribution: it is
not necessarily isotropic. On the other hand, when a shear stress
$\taut$ is applied during the liquid-solid transition, it is
assumed that some of the doublets change their orientation so that
Eqs.~(\ref{eq:cauchy}) hold with $\sigma_{xy} = \taut$, while all
the other quantities describing the morphology of the suspension
at the particle scale ($\lm$, $\lp$, $\fm$, $\fp$, $\km$, $\kp$)
remain unchanged.

In our picture, the distribution of orientation frozen at the
liquid/solid transition is modified if $\taut\neq0$. We choose to
describe the modified distributions of orientation for the
doublets by:
\begin{equation}
  \label{eq:newp}
  \begin{array}{lll}
    \pp & = \ppo - {\displaystyle \frac{\dnp}{\pi}} +
    \dnp \delta (\theta - \pi/4) \\
  \\
    \pm & = \pmo - {\displaystyle \frac{\dnm}{\pi}} +
    \dnm \delta (\theta + \pi/4)
  \end{array}
\end{equation}
where $\dnp$ (resp. $\dnm$) denotes the number of attractive
(resp. repulsive) doublets that change their orientation with
respect to the distribution defined by $\ppo$ (resp. $\pmo$), and
$\delta (\theta)$ is the Dirac distribution. The new distribution
of attractive doublets is obtained from the original one by
assuming that $\dnp$ doublets, initially uniformly distributed in
space, rotate from their original position to the position defined
by $\theta = \pi/4$. The same phenomena occur for $\dnm$ repulsive
doublets rotating from their original orientation to the position
$\theta=-\pi/4$. These preferential directions are chosen
consistently with what is observed under shear
\cite{mathis1988,parsi1987,Foss2000}.

Putting Eqs.~(\ref{eq:newp}) into Eqs.~(\ref{eq:cauchy}) yields
the stress tensor value at the liquid-solid transition in this
simple picture:
\begin{equation}
  \label{eq:cauchy-2}
 \sigma_{xx} = \sigma_{yy} = 0 \qquad
 \sigma_{xy} = \taut = {\displaystyle \frac{1}{2 |V|}
 \left(\fp \lp \dnp + \fm \lm \dnm \right)}
\end{equation}
i.e. there is a direct link between the number of doublets that
change their orientation to preferential directions and $\taut$
which is the stress at the liquid/solid transition. On the other
hand, from Eqs.~(\ref{eq:newp}) and (\ref{eq:Ghom}), the elastic
modulus reads:
\begin{equation}
  \label{eq:Ghom-2}
  G = G_0 + \frac{1}{8|V|} \left(\kp (\lp)^2 \dnp + \km (\lm)^2 \dnm \right)
\end{equation}
i.e. the elastic modulus is increased by a quantity proportional
to the number of doublets changing their orientation. From
Eqs.~(\ref{eq:cauchy-2}) and~(\ref{eq:Ghom-2}), it is thus clearly
seen that $\Delta G(\taut)$ is roughly proportional to $\taut$,
and exactly proportional in some particular cases. If e.g.
$\dnp=\dnm$, we have:
\begin{equation}
  \label{eq:final}
  G-G_0 =  \frac{1}{4}\frac{\kp (\lp)^2 + \km (\lm)^2
   }{\fp \lp + \fm \lm} \taut
\end{equation}
where $\frac{\kp (\lp)^2 + \km (\lm)^2
   }{\fp \lp + \fm \lm} >0$. To sum up, within this simple picture of a 2D model colloidal suspension with repulsive and
attractive pair interactions, we have shown that a liquid/solid
transition under a stress $\taut\neq0$ imply a particle pair
change of orientation that yields an increase of the elastic
modulus roughly proportional to $\taut$.

\end{document}